# Mixed Strategies in Combinatorial Agency

**Moshe Babaioff**                                      MOSHE@MICROSOFT.COM
*Microsoft Research - Silicon Valley*
*Mountain View, CA 94043 USA*

**Michal Feldman**                                      MFELDMAN@HUJI.AC.IL
*School of Business Administration*
*and Center for the Study of Rationality,*
*Hebrew University of Jerusalem,*
*Jerusalem, Israel*

**Noam Nisan**                                          NOAM@CS.HUJI.AC.IL
*School of Computer Science,*
*The Hebrew University of Jerusalem,*
*Jerusalem, Israel*

## Abstract

In many multiagent domains a set of agents exert effort towards a joint outcome, yet the individual effort levels cannot be easily observed. A typical example for such a scenario is routing in communication networks, where the sender can only observe whether the packet reached its destination, but often has no information about the actions of the intermediate routers, which influences the final outcome. We study a setting where a principal needs to motivate a team of agents whose combination of hidden efforts stochastically determines an outcome. In a companion paper we devise and study a basic "combinatorial agency" model for this setting, where the principal is restricted to inducing a *pure* Nash equilibrium. Here we study various implications of this restriction. First, we show that, in contrast to the case of observable efforts, inducing a mixed-strategies equilibrium may be beneficial for the principal. Second, we present a sufficient condition for technologies for which no gain can be generated. Third, we bound the principal's gain for various families of technologies. Finally, we study the robustness of mixed equilibria to coalitional deviations and the computational hardness of the optimal mixed equilibria.

## 1. Introduction

In this paper we study Combinatorial Agency with Mixed Strategies, this section reviews some background on Combinatorial Agency with pure strategies and then present our results for mixed strategies.

### 1.1 Background: Combinatorial Agency

The well studied principal-agent problem deals with how a "principal" can motivate a rational "agent" to exert costly effort towards the welfare of the principal. The difficulty in this model is that the agent's action (i.e. whether he exerts effort or not) is invisible to the principal and only the final outcome, which is probabilistic and also influenced by other factors, is visible[1]. This problem is well studied in many contexts in classical economic theory and we refer the readers to introductory texts on economic theory such as the work of Mass-Colell, Whinston, and Green (1995), Chapter 14. In these settings, a properly designed contract, in which the payments are contingent upon the final outcome, can influence a rational agent to exert the required effort.

In many multiagent settings, however, a set of agents work together towards a joint outcome. Handling *combinations* of agents rather than a single agent is the focus of the work by Babaioff, Feldman, and Nisan (2006a). While much work was previously done on motivating teams of agents (e.g., Holmstrom, 1982; Strausz, 1996), our emphasis is on dealing with the complex combinatorial structure of dependencies between agents' actions. In the general case, each combination of efforts exerted by the $n$ different agents may result in a different expected gain for the principal. The general question asks, given an exact specification of the expected utility of the principal for each combination of agents' actions, which conditional payments should the principal offer to which agents as to maximize his net utility?

We view this problem of hidden actions in computational settings as a complementary problem to the problem of hidden information that is the heart of the field of Algorithmic Mechanism Design (Nisan, Roughgarden, Tardos, & Vazirani, 2007; Nisan & Ronen, 2001). In recent years, computer science and artificial intelligence have showed a lot of interest in algorithmic mechanism design. In particular, they imported concepts from game theory and mechanism design for solving problems that arise in artificial intelligence application domains, such as computer networks with routers as autonomous software agents.

Communication networks serve as a typical application to our setting. Since many computer networks (such as the Internet and mobile ad-hoc networks) are used and administered by multiple entities with different economic interests, their performance is determined by the actions among the various interacting self-interested parties. Thus, taking into account the economic and strategic considerations together with the technical ones may be crucial in such settings. Indeed, recent years have seen a flurry of research employing game theoretic models and analysis for better understanding the effect of strategic considerations on network design and performance.

An example that was discussed in the work of Feldman, Chuang, Stoica, and Shenker (2007) is Quality of Service routing in a network: every intermediate link or router may exert a different amount of "effort" (priority, bandwidth, etc.) when attempting to forward a packet of information. While the final outcome of whether a packet reached its destination is clearly visible, it is rarely feasible to monitor the exact amount of effort exerted by each intermediate link – how can we ensure that they really do exert the appropriate amount of effort? For example, in Internet routing, IP routers may delay or drop packets, and in mobile ad hoc networks, devices may strategically drop packets to conserve their constrained energy resources. Aside from forwarding decisions, which are done in a sequential manner, some "effort" decisions take place prior to the actual packet transmission, and are done in a simultaneous manner. There are many examples for such decisions, among them are the quality of the hardware, appropriate tuning of the routers, and more. Our focus is on these a-priori effort decisions, since they are crucial to the quality of the transmission, and it is harder to detect agents who shirk with respect to these matters.

---

1. "Invisible" here is meant in a wide sense that includes "not precisely measurable", "costly to determine", or "non-contractible" (meaning that it can not be upheld in "a court of law").

In the general model presented in the work of Babaioff et al. (2006a), each of $n$ agents has a set of possible *actions*, the combination of actions by the players results in some *outcome*, where this happens probabilistically. The main part of the specification of a problem in this model is a function (the "technology") that specifies this distribution for each $n$-tuple of agents' actions. Additionally, the problem specifies the principal's utility for each possible outcome, and for each agent, the agent's cost for each possible action. The principal motivates the agents by offering to each of them a *contract* that specifies a payment for each possible outcome of the whole project. Key here is that the actions of the players are non-observable ("hidden-actions") and thus the contract cannot make the payments directly contingent on the actions of the players, but rather only on the outcome of the whole project.

Given a set of contracts, each agent optimizes his own utility; i.e., chooses the action that maximizes his expected payment minus the cost of the action. Since the outcome depends on the actions of all players together, the agents are put in a game here and are assumed to reach a Nash Equilibrium (NE). The principal's problem is that of designing the *optimal contract*: i.e. the vector of contracts to the different agents that induce an equilibrium that will optimize his expected utility from the outcome minus his expected total payment. The main difficulty is that of determining the required Nash equilibrium point.

Our interest in this paper, as in the work of Babaioff et al. (2006a), is focused on the binary case: each agent has only two possible actions "exert effort" and "shirk" and there are only two possible outcomes "success" and "failure". Our motivating examples come from the following more restricted and concrete "structured" subclass of problem instances: Every agent $i$ performs a subtask which succeeds with a low probability $\gamma_i$ if the agent does not exert effort and with a higher probability $\delta_i > \gamma_i$, if the agent does exert effort. The whole project succeeds as a deterministic Boolean function of the success of the subtasks. For example, the Boolean "AND" and "OR" functions represent the respective cases where the agents are complementary (i.e., where the project succeeds if and only if all the agents succeed) or substitutive (i.e., where the project succeeds if and only if at least one of the agents succeeds). Yet, a more restricted subclass of problem instances are those technologies that can be represented by "read-once" networks with two specified source and sink nodes, in which every edge is labeled by a single agent, and the project succeeds (e.g., a packet of information reaches the destination) if there is a successful path between the source and the sink nodes.

## 1.2 This Paper: Mixed Equilibria

The focus in the work by Babaioff et al. (2006a) was on the notion of Nash-equilibrium in pure strategies: we did not allow the principal to attempt inducing an equilibrium where agents have mixed strategies over their actions. In the observable-actions case (where the principal can condition the payments on the agents' individual actions) the restriction to pure strategies is without loss of generality: mixed actions can never help since they simply provide a convex combination of what would be obtained by pure actions.

Yet, surprisingly, we show this is not the case for the hidden-actions case which we are studying: in some cases, a Mixed-Nash equilibrium can provide better expected utility to the principal than what he can obtain by equilibrium in pure strategies. In particular, this already happens in the case of two substitutive agents with a certain (quite restricted) range of parameters (see Section 3).

While inducing mixed strategy equilibria might be beneficial for the principal, mixed Nash equilibrium is a much weaker solution concept than pure Nash equilibrium, as was already observed by Harsanyi (1973). As opposed to Nash equilibria in pure strategies, the guarantees that one obtains are only in expectation. In addition, any player can deviate from his equilibrium strategy without lowering his expected payoff even if he expects all other players to stick to their equilibrium strategies. Moreover, best-response dynamics converge to pure profiles, and there is no natural dynamics leading to a mixed Nash equilibrium. As a result, if the principal cannot gain much by inducing a Nash equilibrium in mixed strategies, he might not be willing to tolerate the instability of this notion. Our main goal is to quantify the principal's gain from inducing mixed equilibrium, rather than pure. To do that, we analyze the worst ratio (over all principal's values) between the principal's optimal utility with mixed equilibrium, and his optimal utility with pure equilibrium. We term this ratio "the price of purity" (POP) of the instance under study.

The price of purity is at least 1 by definition, and the larger it is, the more the principal can gain by inducing a mixed equilibrium compared to a pure one. We prove that for super-modular technologies (e.g. technologies with "increasing returns to scale") which contains in particular the $AND$ Boolean function, the price of purity is trivial (i.e., $POP = 1$). Moreover, we show that for any other Boolean function, there is an assignment of the parameters (agents' individual success probabilities) for which the obtained structured technology has non trivial POP (i.e., $POP > 1$). (Section 4).

While the price of purity may be strictly greater than 1, we obtain quite a large number of results bounding this ratio (Section 5). These bounds range from a linear bound for very general families of technologies (e.g., $POP \leq n$ for any anonymous or sub-modular technology) to constant bounds for some restricted cases (e.g., $POP \leq 1.154...$ for a family of anonymous OR technologies, and $POP \leq 2$ for any technology with 2 agents).

Additionally, we study some other properties of mixed equilibrium. We show that mixed Nash equilibria are more delicate than pure ones. In particular, we show that unlike the pure case, in which the optimal contract is also a "strong equilibrium" (Aumann, 1959) (i.e., resilient to deviations by coalitions), an optimal mixed contract (in which at least two agents truly mix) never satisfies the requirements of a strong equilibrium (Section 6).

Finally, we study the computational hardness of the optimal mixed Nash equilibrium, and show that the hardness results from the pure case hold for the mixed case as well (Section 7).

## 2. Model and Preliminaries

We focus on the simple "binary action, binary outcome" scenario where each agent has two possible actions ("exert effort" or "shirk") and there are two possible outcomes ("failure", "success"). We begin by presenting the model with pure actions (which is a generalization of the model of Winter, 2004), and then move to the mixed case. A principal employs a set of agents $N$ of size $n$. Each agent $i \in N$ has a set of two possible actions $A_i = \{0, 1\}$ (binary action), the low effort action (0) has a cost of 0 ($c_i(0) = 0$), while the high effort action (1) as a cost of $c_i > 0$ ($c_i(1) = c_i$). The played profile of actions determine, in a probabilistic way, a "contractible" outcome, $o \in \{0, 1\}$, where the outcomes 0 and 1 denote project failure and success, respectively (binary-outcome). The outcome is determined according to a success function $t : A_1 \times \ldots \times A_n \to [0, 1]$, where $t(a_1, \ldots, a_n)$ denotes the probability of project success where players play with the action profile $a = (a_1, \ldots, a_n) \in A_1 \times \ldots \times A_n = A$. We use the notation $(t, \vec{c})$ to denote a technology (a success function and a vector of costs, one for each agent). We assume that everything but the effort of the agents is common knowledge.

The principal's value of a successful project is given by a scalar $v > 0$, where he gains no value from a project failure. In this hidden-actions model the actions of the players are invisible, but the final outcome is visible to him and to others, and he may design enforceable contracts based on this outcome. We assume that the principal can pay the agents but not fine them (known as the *limited liability* constraint). The contract to agent $i$ is thus given by a scalar value $p_i \geq 0$ that denotes the payment that $i$ gets in case of project success. If the project fails, the agent gets no money (this is in contrast to the "observable-actions" model in which payment to an agent can be contingent on his action). The contracts to all the agents public, all agents know them before making their effort decisions.

Given this setting, the agents have been put in a game, where the utility of agent $i$ under the profile of actions $a = (a_1, \ldots, a_n) \in A$ is given by $u_i(a) = p_i \cdot t(a) - c_i(a_i)$. As usual, we denote by $a_{-i} \in A_{-i}$ the $(n-1)$-dimensional vector of the actions of all agents excluding agent $i$. i.e., $a_{-i} = (a_1, \ldots, a_{i-1}, a_{i+1}, \ldots, a_n)$. The agents will be assumed to reach Nash equilibrium, if such an equilibrium exists. The principal's problem (which is our problem in this paper) is how to design the contracts $p_i$ as to maximize his own expected utility $u(a, v) = t(a) \cdot (v - \sum_{i \in N} p_i)$, where the actions $a_1, \ldots, a_n$ are at Nash-equilibrium. In the case of multiple Nash equilibria, in our model we let the principal choose the desired one, and "suggest" it to the agents, thus focusing on the "best" Nash equilibrium.[2]

As we wish to concentrate on motivating agents, rather than on the coordination between agents, we assume that more effort by an agent always leads to a better probability of success. Formally, $\forall i \in N, \forall a_{-i} \in A_{-i}$ we have that $t(1, a_{-i}) > t(0, a_{-i})$. We also assume that $t(a) > 0$ for any $a \in A$.

We next consider the extended game in which an agent can mix between exerting effort and shirking (randomize over the two possible pure actions). Let $q_i$ denote the probability that agent $i$ exerts effort, and let $q_{-i}$ denote the $(n-1)$-dimensional vector of investment probabilities of all agents except for agent $i$. We can extend the definition of the success function $t$ to the range of mixed strategies, by taking the expectation.

$$t(q_1, \ldots, q_n) = \sum_{a \in \{0,1\}^n} (\prod_{i=1}^{n} q_i^{a_i} \cdot (1 - q_i)^{(1-a_i)}) t(a_1, \ldots, a_n)$$

---

2. While in the pure case (Babaioff, Feldman, & Nisan, 2006b), the best Nash equilibrium is also a strong equilibrium, this is not the case in the more delicate mixed case (see Section 6). Other variants of NE exist. One variant, which is similar in spirit to "strong implementation" in mechanism design, would be to take the worst Nash equilibrium, or even, stronger yet, to require that only a single equilibrium exists (as in the work of Winter, 2004).

Note that for any agent $i$ and any $(q_i, q_{-i})$ it holds that $t(q_i, q_{-i}) = q_i \cdot t(1, q_{-i}) + (1 - q_i) \cdot t(0, q_{-i})$. A mixed equilibrium profile in which at least one agent mixes with probability $p_i \in [0, 1]$ is called a *non-degenerate* mixed equilibrium.

In pure strategies, the marginal contribution of agent $i$, given $a_{-i} \in A_{-i}$, is defined to be: $\Delta_i(a_{-i}) = t(1, a_{-i}) - t(0, a_{-i})$. For the mixed case we define the marginal contribution of agent $i$, given $q_{-i}$ to be: $\Delta_i(q_{-i}) = t(1, q_{-i}) - t(0, q_{-i})$. Since $t$ is monotone, $\Delta_i$ is a positive function.

We next characterize what payment can result in an agent mixing between exerting effort and shirking.

**Claim 2.1** *Agent $i$'s best response is to mix between exerting effort and shirking with probability $q_i \in (0, 1)$ only if he is indifferent between $a_i = 1$ and $a_i = 0$. Thus, given a profile of strategies $q_{-i}$, agent $i$ mixes only if:*

$$p_i = \frac{c_i}{\Delta_i(q_{-i})} = \frac{c_i}{t(1, q_{-i}) - t(0, q_{-i})}$$

*which is the payment that makes him indifferent between exerting effort and shirking. The expected utility of agent $i$, who exerts effort with probability $q_i$ is: $u_i(q) = c_i \cdot \left( \frac{t(q)}{\Delta_i(q_{-i})} - q_i \right)$.*

*Proof:* Recall that $u_i(q) = t(q) \cdot p_i - q_i \cdot c_i$, thus $u_i(q) = q_i \cdot u_i(1, q_{-i}) + (1 - q_i) \cdot u_i(0, q_{-i})$. Since $i$ maximizes his utility, if $q_i \in (0, 1)$, it must be the case that $u_i(1, q_{-i}) = u_i(0, q_{-i})$. Solving for $p_i$ we get that $p_i = \frac{c_i}{\Delta_i(q_{-i})}$. □

A profile of mixed strategies $q = (q_1, \ldots, q_n)$ is a *Mixed Nash equilibrium* if for any agent $i$, $q_i$ is agent $i$'s best response, given $q_{-i}$.

The principal's expected utility under the mixed Nash profile $q$ is given by $u(q, v) = (v - P) \cdot t(q)$, where $P$ is the total payment in case of success, given by $P = \sum_{i|q_i > 0} \frac{c_i}{\Delta_i(q_{-i})}$. An *optimal mixed contract* for the principal is an equilibrium mixed strategy profile $q^*(v)$ that maximizes the principal's utility at the value $v$. In Babaioff et al. (2006a) we show a similar characterization of optimal pure contract $a \in A$. An agent that exerts effort is paid $\frac{c_i}{\Delta_i(a_{-i})}$, and the utilities are the same as the above, when given the pure profile. In the pure Nash case, given a value $v$, an *optimal pure contract* for the principal is a set of agents $S^*(v)$ that exert effort in equilibrium, and this set maximizes the principal's utility at the value $v$.

A simple but crucial observation, generalizing a similar one in the work of Babaioff et al. (2006a) for the pure Nash case, shows that the optimal mixed contract exhibits some monotonicity properties in the value.

**Lemma 2.2** (**Monotonicity lemma**): *For any technology $(t, \vec{c})$ the expected utility of the principal at the optimal mixed contract, the success probability of the optimal mixed contract, and the expected payment of the optimal mixed contract, are all monotonically non-decreasing with the value.*

The proof is postponed to Appendix A, and it also shows that the same monotonicity also holds in the observable-actions case. Additionally, the lemma holds in more general settings, where each agent has an arbitrary action set (not restricted to the binary-actions model considered here).

We wish to quantify the gain by inducing mixed Nash equilibrium, over inducing pure Nash. We define the *price of purity* as the worse ratio (over $v$) between the maximum utilities that are obtained in mixed and pure strategies.

**Definition 2.3** *The price of purity $POP(t, \vec{c})$ of a technology $(t, \vec{c})$ is defined as the worse ratio, over $v$, between the principal's optimal utility in the mixed case and his optimal utility in the pure case. Formally,*

$$POP(t, \vec{c}) = Sup_{v>0} \frac{t(q^*(v)) \left( v - \sum_{i|q_i^*(v)>0} \frac{c_i}{\Delta_i(q_{-i}^*(v))} \right)}{t(S^*(v)) \left( v - \sum_{i \in S^*(v)} \frac{c_i}{\Delta_i(a_{-i})} \right)}$$

*where $S^*(v)$ denotes an optimal pure contract and $q^*(v)$ denotes an optimal mixed contract, for the value $v$.*

The price of purity is at least 1, and may be greater than 1, as we later show. Additionally, it is obtained at some value that is a transition point of the pure case (a point in which the principal is indifferent between two optimal pure contracts).

**Lemma 2.4** *For any technology $(t, \vec{c})$, the price of purity is obtained at a finite $v$ that is a transition point between two optimal pure contracts.*

## 2.1 Structured Technology Functions

In order to be more concrete, we next present technology functions whose structure can be described easily as being derived from independent agent tasks – we call these *structured technology functions*. This subclass gives us some natural examples of technology functions, and also provides a succinct and natural way to represent technology success functions.

In a structured technology function, each individual succeeds or fails in his own "task" independently. The project's success or failure deterministically depends, maybe in a complex way, on the set of successful sub-tasks. Thus we will assume a monotone Boolean function $f : \{0, 1\}^n \rightarrow \{0, 1\}$ which indicates whether the project succeeds as a function of the success of the $n$ agents' tasks.

A structured technology function $t$ is defined by $t(a_1, \ldots, a_n)$ being the probability that $f(x_1, \ldots, x_n) = 1$ where the bits $x_1, \ldots, x_n$ are chosen according to the following distribution: if $a_i = 0$ then $x_i = 1$ with probability $\gamma_i \in [0, 1)$ (and $x_i = 0$ with probability $1 - \gamma_i$); otherwise, i.e. if $a_i = 1$, then $x_i = 1$ with probability $\delta_i > \gamma_i$ (and $x_i = 0$ with probability $1 - \delta_i$). Thus, a structured technology is defined by $n$, $f$ and the parameters $\{\delta_i, \gamma_i\}_{i \in N}$.

Let us consider two simple structured technology functions, "AND" and "OR". First consider the "AND" technology: $f(x_1, \ldots, x_n)$ is the logical conjunction of $x_i$ ($f(x) = \bigwedge_{i \in N} x_i$). Thus the project succeeds only if all agents succeed in their tasks. This is shown graphically as a read-once network in Figure 1(a). For this technology, the probability of success is the product of the individual success probabilities. Agent $i$ succeeds with probability $\delta_i^{a_i} \cdot \gamma_i^{1-a_i}$, thus $t(a) = \prod_{i \in N} \delta_i^{a_i} \cdot \gamma_i^{1-a_i}$.

Next, consider the "OR" technology: $f(x_1, \ldots, x_n)$ is the logical disjunction of $x_i$ ($f(x) = \bigvee_{i \in N} x_i$). Thus the project succeeds if at least one of the agents succeed in their tasks. This is shown graphically as a read-once network in Figure 1(b). For this technology, the probability of success is 1 minus the probability that all of them fail. Agent $i$ fails with probability $(1 - \delta_i)^{a_i} \cdot (1 - \gamma_i)^{1-a_i}$, thus $t(a) = 1 - \prod_{i \in N}(1 - \delta_i)^{a_i} \cdot (1 - \gamma_i)^{1-a_i}$.

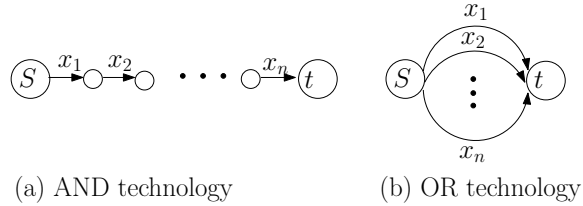

(a) AND technology  (b) OR technology

Figure 1: AND and OR technologies. In AND (a), the project is successful if a packet is routed along a linear path (where each agent controls an edge), and in OR (b), the project is successful if a packet is routed at least along one edge.

These are just two simple examples. One can consider other more interesting examples as the Majority function (the project succeed if the majority of the agents are successful), or the OR-Of-ANDs technology, which is a disjunction over conjunctions (several teams, the project succeed if all the agents in any one of the teams are successful). For additional examples see the work of Babaioff et al. (2006a).

A success function $t$ is called *anonymous* if it is symmetric with respect to the players. I.e. $t(a_1, \ldots, a_n)$ depends only on $\sum_i a_i$. For example, in an anonymous OR technology there are parameters $1 > \delta > \gamma > 0$ such that each agent $i$ succeed with probability $\gamma$ with no effort, and with probability $\delta > \gamma$ with effort. If $m$ agents exert effort, the success probability is $1 - (1-\delta)^m \cdot (1-\gamma)^{n-m}$.

A technology has *identical costs* if there exists a $c$ such that for any agent $i$, $c_i = c$. As in the case of identical costs the POP is independent of $c$, we use $POP(t)$ to denote the POP for technology $t$ with identical costs. We abuse notation and denote a technology with identical costs by its success function $t$. Throughout the paper, unless explicitly stated otherwise, we assume identical costs. A technology $t$ with identical costs is *anonymous* if $t$ is anonymous.

## 3. Example: Mixed Nash Outperforms Pure Nash!

If the actions are observable (henceforth, the observable-actions case), then an agent that exerts effort is paid exactly his cost, and the principal's utility equals the social welfare. In this case, the social welfare in mixed strategies is a convex combination of the social welfare in pure strategies; thus, it is clear that the optimal utility is always obtained in pure strategies. However, surprisingly enough, in the hidden-actions case, the principal might gain higher utility when mixed strategies are allowed. This is demonstrated in the following example:

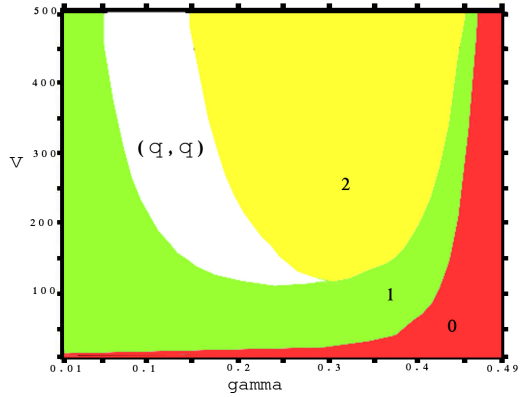

Figure 2: Optimal mixed contracts in *OR* technologies with 2 agents. The areas indicated by "0", "1", and "2" correspond to areas where it is optimal that 0, 1, or 2 agents, respectively, exert effort with probability 1. The white area corresponds to both agents exert effort with the same non-trivial probability, $q$. For any fixed $\gamma$, $q$ increases in $v$.

**Example 3.1** *Consider an anonymous OR technology with two agents, where $c = 1$, $\gamma = \gamma_1 = \gamma_2 = 1 - \delta_1 = 1 - \delta_2 = 0.09$ and $v = 348$. It holds that $t(0,0) = 1 - (1 - \gamma)^2 = 0.172$, $t(0,1) = t(1,0) = 1 - (1 - \gamma)(1 - \delta) = 0.9181$, and $t(1,1) = 1 - (1 - \delta)^2 = 0.992$.*

*Consider the mixed strategy $q_1 = q_2 = 0.92$. It holds that: $t(0, 0.92) = 0.08 \cdot t(0,0) + 0.92 \cdot t(0,1) = 0.858$, $t(1, 0.92) = 0.92 \cdot t(1,1) + 0.08 \cdot t(1,0) = 0.986$, and $t(0.92, 0.92) = 0.08^2 \cdot t(0,0) + 0.08 \cdot 0.92 \cdot t(0,1) \cdot 2 + 0.92^2 \cdot t(1,1) = 0.976$. The payment to each player under a successful project is $p_i(0.92, 0.92) = \frac{1}{t(1,0.92) - t(0,0.92)} = 7.837$, thus the principal's utility under the mixed strategies $q_1 = q_2 = 0.92$ and $v = 348$ is $u((0.92, 0.92), 348) = t(0.92, 0.92) \cdot (348 - 2 \cdot 7.837) = 324.279$.*

*While the principal's utility under the mixed profile $q_1 = q_2 = 0.92$ is $324.279$, the optimal contract with pure strategies is obtained when both agents exert effort and achieves a utility of $318.3$. This implies that by moving from pure strategies to mix strategies, one gains at least $324.27/318.3 > 1.0187$ factor improvement (which is approximately $1.8\%$).*

A worse ratio exists for the more general case (in which it does not necessarily hold that $\delta = 1 - \gamma$) of $\gamma = 0.0001$, $\delta = 0.9$ and $v = 233$. For this case we get that the optimal pure contract is with one agent, gives utility of $208.7$, while the mixed contract $q_1 = q_2 = 0.92$ gives utility of $213.569$, and the ratio is at least $1.0233$ (approximately $2.3\%$).

To complete the example, Diagram 2 presents the optimal contract for *OR* of 2 agents, as a function of $\gamma$ (when $\delta = 1 - \gamma$) and $v$. It shows that for some parameters of $\gamma$ and $v$, the optimal contract is obtained when both agents exert effort with equal probabilities.

The following lemma (proved in Appendix A.1) shows that optimal mixed contracts in any anonymous OR technology (with $n$ agents) have this specific structure. That is, all agents that do not shirk, mix with exactly the same probability.

**Lemma 3.2** *For any anonymous OR technology (any $\delta > \gamma, c, n$) and value $v$, either the optimal mixed contract is a pure contract, or, in the optimal mixed contract $k \in \{2, \ldots n\}$ agents exert effort with equal probabilities $q_1 = \ldots = q_k \in (0, 1)$, and the rest of the agents exert no effort.*

## 4. When is Pure Nash Good Enough?

Next, we identify a class of technologies for which the price of purity is 1; that is, the principal cannot improve his utility by moving from pure Nash equilibrium to mixed Nash equilibrium. These are technologies for which the marginal contribution of any agent is non-decreasing in the effort of the other agents. Formally, for two pure action profiles $a, b \in A$ we denote $b \succeq a$ if for all $j$, $b_j \succeq_j a_j$ (effort $b_j$ is at least as high as the effort $a_j$).

**Definition 4.1** *A technology success function $t$ exhibits (weakly)* **increasing returns to scale (IRS)**[3] *if for every $i$, and every pure profiles $b \succeq a$*

$$t(b_i, b_{-i}) - t(a_i, b_{-i}) \geq t(b_i, a_{-i}) - t(a_i, a_{-i})$$

Any AND technology exhibits IRS (Winter, 2004; Babaioff et al., 2006a). For IRS technologies we show that $POP = 1$.

**Theorem 4.2** *Assume that $t$ is super-modular. For any cost vector $\vec{c}$, $POP(t, \vec{c}) = 1$. Moreover, a non-degenerate mixed contract is never optimal.*

*Proof:* For a mixed profile $q = (q_1, q_2, \ldots, q_n)$, let $S(q)$ be the support of $q$, that is, $i \in S(q)$ if and only if $q_i > 0$, and for any agent $i \in S(q)$ let $S_{-i} = S(q) \setminus \{i\}$ be the support of $q$ excluding $i$. Similarly, for a pure profile $a = (a_1, a_2, \ldots, a_n)$ let $S(a)$ be the support $a$. Under the mixed profile $q$, agent $i \in S(q)$ is being paid $p_i(q_{-i}) = \frac{c_i}{t(1, q_{-i}) - t(0, q_{-i})}$. Similarly, under the pure profile $a$, agent $i \in S(a)$ is being paid $p_i(S(a) \setminus \{i\}) = p_i(a_{-i}) = \frac{c_i}{t(S(a)) - t(S(a) \setminus \{i\})}$, where $t(T)$ is the success probability when $a_j = 1$ for $j \in T$, and $a_j = 0$ for $j \notin T$. We also denote $\Delta_i(T) = t(T) - t(T \setminus \{i\})$.

We show that if $q$ is a non-degenerate mixed profile (i.e., at least one agent in $q$ exerts effort with probability $q_i \in (0, 1)$), the profile in which each agent in $S(q)$ exerts effort with probability 1 yields a higher utility to the principal.

By Lemma 5.3 (see Section 5), it holds that $p_i(q_{-i}) \geq min_{T \subseteq S_{-i}} p_i(T)$, where $p_i(T) = \frac{c_i}{\Delta_i(T)}$. But if $t$ exhibits IRS, then $\Delta_i(T)$ is an increasing function by definition (see Section 4), therefore $min_{T \subseteq S_{-i}} p_i(T) = p_i(S_{-i})$. Therefore it holds that for any $i \in S(q)$, $p_i(q_{-i}) \geq p_i(S_{-i})$, thus:

$$\sum_{i \in S(q)} p_i(q_{-i}) \geq \sum_{i \in S(q)} p_i(S_{-i})$$

---

3. Note that $t$ exhibits IRS if and only if it is super-modular.

In addition, due to the monotonicity of $t$, it holds that $t(q) < t(S(q))$. Therefore,

$$
\begin{aligned}
u(q, v) &= t(q) \left( v - \sum_{i \in S(q)} p_i(q_{-i}) \right) \\
&< t(S(q)) \left( v - \sum_{i \in S(q)} p_i(q_{-i}) \right) \\
&\leq t(S(q)) \left( v - \sum_{i \in S(q)} p_i(S_{-i}) \right) \\
&= u(S(q), v)
\end{aligned}
$$

where $u(S(q), v)$ is the principal's utility under the pure profile in which all the agents in $S(q)$ exert effort with probability 1, and the rest exert no effort. □

We show that $AND$ (on some subset of bits) is the only function such that any structured technology based on this function exhibits IRS, that is, this is the only function such that for any choices of parameters (any $n$ and any $\{\delta_i, \gamma_i\}_{i \in N}$), the structured technology exhibits IRS. For any other Boolean function, there is an assignment for the parameters such that the created structured technology is essentially OR over 2 inputs (Lemma B.1 in Appendix B), thus it has non-trivial POP (recall Example 3.1). For the proof of the following theorem see Appendix B.

**Theorem 4.3** *Let $f$ be any monotone Boolean function with $n \geq 2$ inputs, that is not constant and not a conjunction of some subset of the input bits. Then there exist parameters $\{\gamma_i, \delta_i\}_{i=1}^n$ such that the POP of the structured technology with the above parameters (and identical cost $c = 1$) is greater than $1.0233$.*

Thus, our goal now is to give upper bounds on the POP for various technologies.

## 5. Quantifying the Gain by Mixing

In this section we present bounds on the price of purity for general technologies, following by bounds for the special case of OR technology.

### 5.1 POP for General Technologies

We first show that the POP can be bounded by the principal's price of unaccountability (Babaioff et al., 2006b), whose definition follows.

**Definition 5.1** *The principal's price of unaccountability $POU_P(t, \vec{c})$ of a technology $(t, \vec{c})$ is defined as the worst ratio (over $v$) between the principal's utility in the observable-actions case and the hidden-actions case:*

$$
POU_P(t, \vec{c}) = Sup_{v > 0} \frac{t(S_{oa}^*(v)) \cdot v - \sum_{i \in S_{oa}^*(v)} c_i}{t(S^*(v)) \cdot v - \sum_{i \in S^*(v)} \frac{c_i}{\Delta_i(a_{-i})}}
$$

*where $S_{oa}^*(v)$ is the optimal pure contract in the observable-actions case, and $S^*(v)$ is the optimal pure contract in the hidden-actions case.*

**Theorem 5.2** *For any technology $t$ it holds that $POU_P(t) \geq POP(t)$.*

*Proof*: Both $POU_P(t)$ and $POP(t)$ are defined as supremum over utilities ratio for a given value $v$. We present a bound for any $v$, thus it holds for the supremum. The denominator in both case is the same: it is the optimal utility of the principal in the hidden-actions case with pure strategies. The numerator in the POP is the optimal principal utility in the hidden-actions case with mixed strategies. Obviously, this is at most the optimal principal utility in the observable-actions case with mixed strategies. It has already been observed that in the observable-actions case mixed strategies cannot help the principal (see Section 3), i.e., the principal utility with mixed strategies equals the principal utility with pure strategies. The assertion of the theorem follows by observing that the optimal principal utility with pure strategies in the observable-action case is the numerator of $POU_P$. □

However, this bound is rather weak. To best see this, note that the principal's price of unaccountability for AND might be unbounded (e.g., Babaioff et al., 2006b). Yet, as shown in Section 4.2, $POP(AND) = 1$.

In this section we provide better bounds on technologies with identical costs. We begin by characterizing the payments for a mixed contract. We show that under a mixed profile, each agent in the support of the contract is paid at least the minimal payment to a single agent under a pure profile with the same support, and at most the maximal payment.

For a mixed profile $q = (q_1, q_2, \ldots, q_n)$, let $S(q)$ be the support of $q$, that is, $i \in S(q)$ if and only if $q_i > 0$. Similarly, for a pure profile $a = (a_1, a_2, \ldots, a_n)$ let $S(a)$ be the support $a$. Under the mixed profile $q$, agent $i \in S(q)$ is being paid $p_i(q_{-i}) = \frac{c_i}{t(1,q_{-i})-t(0,q_{-i})}$. Similarly, under the pure profile $a$, agent $i \in S(a)$ is being paid $p_i(S(a) \setminus \{i\}) = p_i(a_{-i}) = \frac{c_i}{t(S(a))-t(S(a)\setminus\{i\})}$, where $t(T)$ is the success probability when $a_j = 1$ for $j \in T$, and $a_j = 0$ for $j \notin T$.

**Lemma 5.3** *For a mixed profile $q = (q_1, q_2, \ldots, q_n)$, and for any agent $i \in S(q)$ let $S_{-i} = S(q) \setminus \{i\}$ be the support of $q$ excluding $i$. It holds that*

$$max_{T \subseteq S_{-i}} \ p_i(T) \geq p_i(q_{-i}) \geq min_{T \subseteq S_{-i}} \ p_i(T)$$

*Proof*: We show that for any agent $i \in S(q)$, the increase in the success probability from him exerting effort when some other players play mixed strategies, is a convex combination of the increases in the success probability when the agents in the support play pure strategies.

Recall that: $t(q_1, \ldots, q_n) = \sum_{a \in \{0,1\}^n} (\prod_{i=1}^n q_i^{a_i} \cdot (1-q_i)^{(1-a_i)}) t(a_1, \ldots, a_n)$.

Let $t'$ be the technology $t$ restricted to the support $S = S(q)$, that is, if $i_1, \ldots, i_S$ are the agents in $S$ then $t'(a_{i_1}, a_{i_2}, \ldots, a_{i_S})$ is defined to be the value of $t$ on $a$, when $a_j = 0$ for any agent $j \notin S$, and $a_j = a_{i_k}$ for $j = i_k \in S$. $t'$ is defined on mixed strategies in the expected way. Thus,

$$
\begin{aligned}
\Delta_i(q_{-i}) &= t(1, q_{-i}) - t(0, q_{-i}) \\
&= t'(1, q_{S_{-i}}) - t'(0, q_{S_{-i}}) \\
&= \sum_{a \in \{0,1\}^{|S|-1}} (\prod_{j \in S_{-i}} q_j^{a_j} \cdot (1-q_j)^{(1-a_j)}) \ t'(1,a) - \sum_{a \in \{0,1\}^{|S|-1}} (\prod_{j \in S_{-i}} q_j^{a_j} \cdot (1-q_j)^{(1-a_j)}) t'(0,a) \\
&= \sum_{a \in \{0,1\}^{|S|-1}} (\prod_{j \in S_{-i}} q_j^{a_j} \cdot (1-q_j)^{(1-a_j)})(t'(1,a) - t'(0,a))
\end{aligned}
$$

We conclude that $\Delta_i(q_{-i})$ is a convex combination of $\Delta_i(b_{-i})$ for $b$ with support $S(b) \subseteq S_{-i}$. Therefore, $min_{T \subseteq S_{-i}} \, (t(\{i\} \cup T) - t(T)) \leq \Delta_i(q_{-i}) \leq max_{T \subseteq S_{-i}} \, (t(\{i\} \cup T) - t(T))$. Thus,

$$
\begin{aligned}
max_{T \subseteq S_{-i}} \, 1/(t(\{i\} \cup T) - t(T)) &= 1/min_{T \subseteq S_{-i}} \, (t(\{i\} \cup T) - t(T)) \\
&\geq 1/\Delta_i(q_{-i}) = p_i(q_{-i}) \\
&\geq 1/max_{T \subseteq S_{-i}} \, (t(\{i\} \cup T) - t(T)) \\
&= min_{T \subseteq S_{-i}} \, 1/(t(\{i\} \cup T) - t(T))
\end{aligned}
$$

$\square$

In what follows, we consider two general families of technologies with $n$ agents: anonymous technologies and technologies that exhibit decreasing returns to scale (DRS). DRS technologies are technologies with decreasing marginal contribution (more effort by others decrease the contribution of an agent). For both families we present a bound of $n$ on the POP.

We begin with a formal definition of DRS technologies.

**Definition 5.4** *A technology success function $t$ exhibits (weakly)* **decreasing returns to scale (DRS)**[4] *if for every $i$, and every $b \succeq a$*

$$t(b_i, b_{-i}) - t(a_i, b_{-i}) \leq t(b_i, a_{-i}) - t(a_i, a_{-i})$$

**Theorem 5.5** *For any anonymous technology or a (non-anonymous) technology that exhibits DRS, it holds that $POP(t) \leq n$.*

For the proof of this theorem as well as the proofs of all claims that appear later in this section, see Appendix C. We also prove a bound on the POP for any technology with 2 agents (even not anonymous), and an improved bound for the anonymous case.

**Theorem 5.6** *For any technology $t$ (even non-anonymous) with 2 agents, it holds that $POP(t) \leq 2$. If $t$ is anonymous then $POP(t) \leq 3/2$.*

We do not provide bounds for non-anonymous technologies, this is left as an open problem for future research. We believe that the linear bound for anonymous and DRS technologies are not tight and we conjecture that there exists a universal constant $C$ that bounds the POP for any technology. Moreover, our simulations seem to indicate that a non-anonymous OR technology with 2 agents yields the highest possible POP. This motivates us to explore the POP for the OR technology in more detail.

## 5.2 POP for the OR Technology

As any OR technology (even non-anonymous) exhibits DRS (see Appendix A.1), this implies a bound of $n$ on the POP of the OR technology. Yet, for anonymous OR technology we present improved bounds on the POP. In particular, if $\gamma = 1 - \delta < 1/2$ we can bound the POP by 1.154....

---

4. Note that $t$ exhibits DRS if and only if it is submodular.

**Theorem 5.7** *For any anonymous OR technology with $n$ agents:*

1. *If $1 > \delta > \gamma > 0$: (a) $POP \leq \frac{1-(1-\delta)^n}{\delta} \leq n - (n-1)\delta$. (b) POP goes to 1 as $n$ goes to $\infty$ (for any fixed $\delta$) or when $\delta$ goes to 1 (for any fixed $n \geq 2$).*

2. *If $\frac{1}{2} > \gamma = 1 - \delta > 0$: (a) $POP \leq \frac{2(3-3\sqrt{3})}{3(\sqrt{3}-2)} (= 1.154..)$. (b) POP goes to 1 as $\gamma$ goes to 0 or as $\gamma$ goes to $\frac{1}{2}$ (for any fixed $n \geq 2$).*

While the bounds for anonymous OR technologies for the case in which $\delta = 1 - \gamma$ are much better than the general bounds, they are still not tight. The highest POP we were able to obtain by simulations was of 1.0233 for $\delta > \gamma$, and 1.0187 for $\delta = 1 - \gamma$ (see Section 3), but deriving the exact bound analytically is left as an open problem.

## 6. The Robustness of Mixed Nash Equilibria

In order to induce an agent $i$ to truly mix between exerting effort and shirking, $p_i$ must be equal exactly to $c_i/\Delta_i(q_{-i})$ (see claim 2.1). Even under an increase of $\epsilon$ in $p_i$, agent $i$ is no longer indifferent between $a_i = 0$ and $a_i = 1$, and the equilibrium falls apart. This is in contrast to the pure case, in which any $p_i \geq \frac{c_i}{\Delta_i(a_{-i})}$ will maintain the required equilibrium. This delicacy exhibits itself through the robustness of the obtained equilibrium to deviations in *coalitions* (as opposed to the unilateral deviations as in Nash). A "strong equilibrium" (Aumann, 1959) requires that no subgroup of players (henceforth *coalition*) can coordinate a joint deviation such that every member of the coalition strictly improves his utility.

**Definition 6.1** *A mixed strategy profile $q \in [0,1]^n$ is a strong equilibrium (SE) if there does not exist any coalition $\Gamma \subseteq N$ and a strategy profile $q'_\Gamma \in \times_{i \in \Gamma}[0,1]$ such that for any $i \in \Gamma$, $u_i(q'_{-\Gamma}, q_\Gamma) > u_i(q)$.*

In the work of Babaioff et al. (2006b) we show that under the payments that induce the pure strategy profile $S^*$ as the best pure Nash equilibrium (i.e., the pure Nash equilibrium that maximizes the principal's utility), $S^*$ is also a *strong equilibrium*. In contrast to the pure case, we next show that any non-degenerate mixed Nash equilibrium $q$ in which there exist at least two agents that truly mix (i.e., $\exists i \neq j$ s.t. $q_i, q_j \in (0,1)$), can never be a strong equilibrium. This is because if the coalition $\Gamma = \{i | q_i \in (0,1)\}$ deviate to $q'_\Gamma$ in which each $i \in \Gamma$ exerts effort with probability 1, each agent $i \in \Gamma$ strictly improves his utility (see proof in Appendix D).

**Theorem 6.2** *If the mixed optimal contract $q$ includes at least two agents that truly mix ($\exists i \neq j$ s.t. $q_i, q_j \in (0,1)$), then $q$ is not a strong equilibrium.*

In any OR technology, for example, it holds that in any non-degenerate mixed equilibrium at least two agents truly mix (see lemma 3.2). Therefore, no non-degenerate contract in the OR technology can be a strong equilibrium.

As generically a mixed Nash contract is not a strong equilibrium while a pure Nash contract always is, if the pricipal wishes to induce a strong Nash equilibrium (e.g., when the agents can coordinate their moves), he can restrict himself to inducing a pure Nash equilibrium, and his loss from doing so is bounded by the POP (see Section 5).

## 7. Algorithmic Aspects

The computational hardness of finding the optimal mixed contract depends on the representation of the technology and how it is being accessed. For a black-box access and for the special case of read-once networks, we generalize our hardness results of the pure case (Babaioff et al., 2006b) to the mixed case. The main open question is whether it is possible to find the optimal mixed contract in polynomial time, given a table representation of the technology (the optimal pure contract can be found in polynomial time in this case). Our generalization theorems follow (see proofs in Appendix E).

**Theorem 7.1** *Given as input a black box for a success function $t$ (when the costs are identical), and a value $v$, the number of queries that is needed, in the worst case, to find the optimal mixed contract is exponential in $n$.*

Even if the technology is a structured technology and further restricted to be the source-pair reliability of a read-once network (see (Babaioff et al., 2006b)), computing the optimal mixed contract is hard.

**Theorem 7.2** *The optimal mixed contract problem for read once networks is #P-hard (under Turing reductions).*

## 8. Conclusions and Open Problems

This paper studies a model in which a principal induces a set of agents to exert effort through individual contracts that are based on the final outcome of the project. The focus of this paper is the question how much the principal can benefit when inducing a Nash equilibrium in mixed strategies instead of being restricted to a pure Nash equilibrium (as was assumed in the original model). We find that while in the case of observable actions mixed equilibria cannot yield the principal a higher utility level than pure ones, this can indeed happen under hidden actions. Yet, whether or not mixed equilibria improve the principal's utility depends on the technology of the project. We give sufficient conditions for technologies in which mixed strategies yield no gain to the principal. Moreover, we provide bounds on the principal's gain for various families of technologies. Finally, we show that an optimal contract in non-degenerated mixed Nash equilibrium is not a strong equilibrium (in contrast to a pure one) and that finding such an optimal contract is computationally challenging.

Our model and results raise several open problems and directions for future work. It would be interesting to study the principal's gain (from mixed strategies) for different families of technologies, such as series-parallel technologies. Additionally, the model can be extended beyond the binary effort level used here. Moreover, our focus was on inducing *some* mixed Nash equilibrium, but that equilibrium might not be unique. One can consider other solution concepts such as a *unique* Nash equilibrium or iterative elimination of dominated strategies. Finally, it might be of interest to study the performance gap between pure and mixed Nash equilibria in domains beyond combinatorial agency.

## 9. Acknowledgments

Michal Feldman is partially supported by the Israel Science Foundation (grant number 1219/09) and by the Leon Recanati Fund of the Jerusalem school of business administration.

## Appendix A. General

**Lemma 2.2** (**Monotonicity lemma**) *For any technology $(t, \vec{c})$ the expected utility of the principal at the optimal mixed contract, the success probability of the optimal mixed contract, and the expected payment of the optimal mixed contract, are all monotonically non-decreasing with the value.*

*Proof:* Suppose the profiles of mixed actions $q^1$ and $q^2$ are optimal for $v_1$ and $v_2 < v_1$, respectively. Let $P^1$ and $P^2$ be the total payment in case a successful project, corresponding to the minimal payments that induce $q^1$ and $q^2$ as Nash equilibria, respectively. The utility is a linear function of the value, $u(a, v) = t(a) \cdot (v - P)$ ($P$ is the total payments in case of successful project). As $q^1$ is optimal at $v_1$, $u(q^1, v_1) \geq u(q^2, v_1)$, and as $t(a) \geq 0$ and $v_1 > v_2$, $u(q^2, v_1) \geq u(q^2, v_2)$. We conclude that $u(q^1, v_1) \geq u(q^2, v_2)$, thus the utility is monotonic non-decreasing in the value.

Next we show that the success probability is monotonic non-decreasing in the value. $q^1$ is optimal at $v_1$, thus:

$$t(q^1) \cdot (v_1 - P^1) \geq t(q^2) \cdot (v_1 - P^2)$$

$q^2$ is optimal at $v_2$, thus:

$$t(q^2) \cdot (v_2 - P^2) \geq t(q^1) \cdot (v_2 - P^1)$$

Summing these two equations, we get that $(t(q^1) - t(q^2)) \cdot (v_1 - v_2) \geq 0$, which implies that if $v_1 > v_2$ then $t(q^1) \geq t(q^2)$.

Finally we show that the expected payment is monotonic non-decreasing in the value. As $q^2$ is optimal at $v_2$ and $t(q^1) \geq t(q^2)$, we observe that:

$$t(q^2) \cdot (v_2 - P^2) \geq t(q^1) \cdot (v_2 - P^1) \geq t(q^2) \cdot (v_2 - P^1)$$

or equivalently, $P^2 \leq P^1$, which is what we wanted to show. $\square$

We note that the above lemma also holds for the case of profiles of pure actions, and for the observable-actions case (by exactly the same arguments).

**Lemma 2.4** *For any technology $(t, \vec{c})$, the price of purity is obtained at a finite $v$ that is a transition point between two optimal pure contracts.*

*Proof:* Clearly for a large enough value $v^*$, the ratio is 1, as in both cases all agents exert maximal effort. For small enough values the principal will choose not to contract with any agent in both cases (and the ratio is 1). This is true as at a value that is smaller than any agent's cost, an optimal contract is to contract with no agent in both cases. Let $\underline{v}$ be the supremum on all values for which the principal will choose not to contract with any agent in both cases. Now, the ratio is a continuous function on the compact range $[\underline{v}, v^*]$, thus its supremum is obtained, for some value that is at most $v^*$.

We have seen that the POP is obtained at some value $v$, we next prove that it is obtained at a transition point of the pure case. If $POP = 1$ the claim clearly holds, thus we should only consider the case that $POP > 1$. Let $v$ be the *maximal value* for which the POP is obtained. Assume in contradiction that $v$ is not a transition point between two optimal pure contracts, and that $a$ and $q$ are optimal for the pure and mixed cases, respectively. As $POP > 1$, $q$ is non-degenerate and $u(q, v) > u(a, v)$. Let $P(a)$ and $P(q)$ denote the total payment in case of success for $a$ and $q$, respectively. We next consider two options.

We first consider the case that $t(a) \geq t(q)$. We show that in this case, the utilities ratio for $v - \epsilon$, for some $\epsilon > 0$ is worse than the utilities ratio for $v$, and we get a contradiction. For $\epsilon > 0$ small enough, the optimal pure contract is still $a$, and $u(a, v - \epsilon) > 0$. Let $q_{-\epsilon}$ be the optimal mixed contract at $v - \epsilon$. It holds that

$$POP \geq \frac{u(q_{-\epsilon}, v - \epsilon)}{u(a, v - \epsilon)} \geq \frac{u(q, v - \epsilon)}{u(a, v - \epsilon)} = \frac{u(q, v) - t(q) \cdot \epsilon}{u(a, v) - t(a) \cdot \epsilon} > \frac{u(q, v)}{u(a, v)}$$

where the last strict inequality is by the following argument.

$$\frac{u(q, v) - t(q) \cdot \epsilon}{u(a, v) - t(a) \cdot \epsilon} > \frac{u(q, v)}{u(a, v)} \quad \Leftrightarrow \quad t(q) \cdot u(a, v) < t(a) \cdot u(q, v) \quad \Leftrightarrow \quad P(q) < P(a)$$

and $P(q) < P(a)$ as $u(q, v) = t(q)(v - P(q)) > t(a)(v - P(a)) = u(a, v)$ and $t(a) \geq t(q)$.

Next we consider the case that $t(q) > t(a)$. If $P(q) < P(a)$, the argument that was presented above shows that the utilities ratio for $v - \epsilon$, for some $\epsilon > 0$, is worse than the utilities ratio for $v$, and we get a contradiction. On the other hand, if $P(q) \geq P(a)$ we show that the utilities ratio for $v + \epsilon$, for some $\epsilon > 0$, is at least as large as the utilities ratio for $v$, in contradiction to $v$ being the maximal value for which the POP is obtained. For $\epsilon > 0$ small enough, the optimal pure contract is still $a$ (as $v$ is not a transition point between pure contracts). Let $q_\epsilon$ be the optimal mixed contract at $v + \epsilon$. It holds that

$$POP \geq \frac{u(q_\epsilon, v + \epsilon)}{u(a, v + \epsilon)} \geq \frac{u(q, v + \epsilon)}{u(a, v + \epsilon)} = \frac{u(q, v) + t(q) \cdot \epsilon}{u(a, v) + t(a) \cdot \epsilon} \geq \frac{u(q, v)}{u(a, v)}$$

where the last inequality is by the following argument.

$$\frac{u(q, v) + t(q) \cdot \epsilon}{u(a, v) + t(a) \cdot \epsilon} \geq \frac{u(q, v)}{u(a, v)} \quad \Leftrightarrow t(q) \cdot u(a, v) \geq t(a) \cdot u(q, v) \quad \Leftrightarrow \quad P(q) \geq P(a)$$

which holds by our assumption. □

The following corollary of Lemma 2.4 will be helpful in finding the POP for technologies with 2 agents.

**Corollary A.1** *Assume that technology $t$ with 2 agents and with identical costs exhibits DRS, then the POP is obtained at the transition point of the pure case, to the contract with both agents.*

*Proof*: By Lemma 2.4 the POP is obtained at a transition point of the pure case. If there is a single transition point, between 0 agents and 2 agents, the claim holds. If contracting with a single agent is sometimes optimal, it must be the case that the single agent that is contracted is the agent with the (possibly weakly) highest success probability (agent $i$ such that $t(\{i\}) \geq t(\{j\})$ where $j \neq i$, which implies that $\Delta_i = t(\{i\}) - t(\emptyset) \geq t(\{j\}) - t(\emptyset) = \Delta_j$). Thus we only need to show that the POP is not obtained at the transition point $v$ between 0 agents and the contract with agent $i$. Assume that $q$ is the optimal mixed contract at $v$, and that $P(q)$ is the total payment in case of success. If $q$ gives the same utility as the contract $\{i\}$, we are done.

Otherwise, $u(q, v) > u(\{i\}, v)$, and by Corollary C.9 it holds that $P(q) \geq \frac{c}{\Delta_1}$, thus $t(q) > t(\{i\})$. This implies that the utilities ratio at the value $v + \epsilon$ for $\epsilon > 0$ small enough is worse than the ratio for $v$ (by the argument presented in Lemma 2.4 for the case that $t(q) > t(a)$). □

## A.1 Analysis of the OR Technology

**Lemma 3.2** *For any anonymous OR technology (any $\delta > \gamma, c, n$) and value $v$, either the optimal mixed contract is a pure contract, or, in the optimal mixed contract $k \in \{2, \ldots n\}$ agents exert effort with equal probabilities $q_1 = \ldots = q_k \in (0, 1)$, and the rest of the agents exert no effort.*

*Proof*: First, observe that it cannot by the case that all agents but one exert no effort, and this single agent mix with probability $0 < q_i < 1$. This is so as the principal would rather change the profile to $q_i = 1$ (pays the same, but gets higher success probability). Suppose by contradiction that a contract that induces a profile $(q_i, q_j, q_{-ij})$ such that $q_i, q_j \in (0, 1]$ and $q_i \neq q_j$ ($q_i > q_j$ without loss of generality) is optimal. For agent $k$, we denote the probability of failure of agent $k$ in his task by $\phi(q_k)$. That is, $\phi(q_k) = 1 - (q_k \delta + (1 - q_k)\gamma) = 1 - \gamma + (\gamma - \delta)q_k = \delta + \beta q_k$ where $\beta = \gamma - \delta$.

We show that for a sufficiently small $\epsilon > 0$, the mixed profile $q' = (q_i - \epsilon, q_j + \epsilon \frac{\phi(q_j)}{\phi(q_i)}, q_{-ij})$ (for $\epsilon$ such that $q' \in [0, 1]$. i.e., $\epsilon < min\{q_i, (1 - q_i)\frac{\phi(q_i)}{\phi(q_j)}\}$, ) obtains a better contract, in contradiction to the optimality of the original contract.

For the *OR* technology, $t(q) = 1 - \prod_{k \in N} \phi(q_k) = 1 - \Phi(q)$, where $\Phi(q) = \prod_{k \in N} \phi(q_k)$. We also denote $\Phi_{-ij}(q) = \prod_{k \neq i,j} \phi(q_k)$. The change in success probability is related to the new product $\phi(q_i - \epsilon) \cdot \phi\left(q_j + \frac{\phi_j}{\phi_i}\epsilon\right)$:

$$
\begin{aligned}
\phi(q_i') \cdot \phi(q_j') &= \phi(q_i - \epsilon) \cdot \phi\left(q_j + \epsilon \frac{\phi(q_j)}{\phi(q_i)}\right) \\
&= (\phi(q_i) - \beta\epsilon) \cdot \left(\phi(q_j) + \epsilon\beta \frac{\phi(q_j)}{\phi(q_i)}\right) \\
&= \phi(q_i)\phi(q_j) - \beta\epsilon\phi(q_j) + \epsilon\beta \frac{\phi(q_j)}{\phi(q_i)}\phi(q_i) - \beta^2\epsilon^2 \frac{\phi(q_j)}{\phi(q_i)} \\
&= \phi(q_i)\phi(q_j) - \beta^2\epsilon^2 \frac{\phi(q_j)}{\phi(q_i)}
\end{aligned}
$$

Therefore the new success probability $t(q')$ has increased by the change:

$$t(q') = t(q_i - \epsilon, q_j + \frac{\phi_j}{\phi_i}\epsilon, q_{-ij})$$

$$= 1 - \phi(q_i - \epsilon) \cdot \phi\left(q_j + \epsilon\frac{\phi(q_j)}{\phi(q_i)}\right) \cdot \Phi_{-ij}(q)$$

$$= 1 - \left(\phi(q_i)\phi(q_j) - \beta^2\epsilon^2\frac{\phi(q_j)}{\phi(q_i)}\right) \cdot \Phi_{-ij}(q)$$

$$= t(q) + \frac{\beta^2\epsilon^2\Phi(q)}{(\phi(q_i))^2} = t(q)\left(1 + \frac{\beta^2\epsilon^2\Phi(q)}{t(q)\cdot(\phi(q_i))^2}\right)$$

We denote $z(\epsilon) = \frac{\beta^2\epsilon^2\Phi(q)}{t(q)\cdot(\phi(q_i))^2}$, thus $t(q') = t(q)\cdot(1 + z(\epsilon))$, where $z(\epsilon) > 0$ for any $\epsilon$.

After showing that the success probability increases, we are left to show that for sufficiently small $\epsilon$, the total payment decreases. The payment to agent $l$ is given by:

$$p_l = \frac{c}{t(1, q_{-l}) - t(0, q_{-l})} = \frac{c}{(\delta - \gamma)\prod_{m \neq l}\phi(q_m)} = \frac{c\cdot\phi(q_l)}{(\delta - \gamma)\cdot t(q)}$$

The change in the payment of agent $k$ is

$$p_k - p'_k = \frac{c\cdot\phi(q_k)}{(\delta - \gamma)\cdot t(q)} - \frac{c\cdot\phi(q'_k)}{(\delta - \gamma)\cdot t(q')}$$

$$= \frac{c}{t(q)\cdot(\delta - \gamma)}\cdot\left(\phi(q_k) - \frac{\phi(q'_k)}{(1 + z(\epsilon))}\right)$$

$$= \frac{c}{t(q)\cdot(\delta - \gamma)\cdot(1 + z(\epsilon))}\cdot\left(\phi(q_k) - \phi(q'_k) + \phi(q_k)\cdot z(\epsilon)\right)$$

$$= W(\epsilon)\cdot\left(\phi(q_k) - \phi(q'_k) + \phi(q_k)\cdot z(\epsilon)\right)$$

for $W(\epsilon) = \frac{c}{t(q)\cdot(\delta - \gamma)\cdot(1 + z(\epsilon))}$.

For agent $k \neq i, j$, as $\phi(q_k) = \phi(q'_k)$ we get $p_k - p'_k = W(\epsilon)\cdot\phi(q_k)\cdot z(\epsilon)$. For agent $i$, as $\phi(q_i) - \phi(q'_i) = \beta\epsilon$ we get $p_i - p'_i = W(\epsilon)\cdot(\beta\epsilon + \phi(q_i)\cdot z(\epsilon))$. For agent $j$, as $\phi(q_j) - \phi(q'_j) = -\beta\epsilon\frac{\phi(q_j)}{\phi(q_i)}$ we get $p_j - p'_j = W(\epsilon)\cdot(-\beta\epsilon\frac{\phi(q_j)}{\phi(q_i)} + \phi(q_j)\cdot z(\epsilon))$.

By summing over all agents we get

$$\sum_{k \in N} p_k - \sum_{k \in N} p'_k = \sum_{k \in N}(p_k - p'_k)$$

$$= (p_i - p'_i) + (p_j - p'_j) + \sum_{k \neq i,j}(p_k - p'_k)$$

$$= W(\epsilon)\cdot\left(\beta\epsilon - \beta\epsilon\frac{\phi(q_j)}{\phi(q_i)} + z(\epsilon)\cdot\sum_{k \in N}\phi(q_k)\right)$$

$$= W(\epsilon)\cdot\left(\beta\epsilon\left(1 - \frac{\phi(q_j)}{\phi(q_i)}\right) + z(\epsilon)\cdot\sum_{k \in N}\phi(q_k)\right)$$

which is positive by the following observations. $W(\epsilon) > 0$ and $z(\epsilon) > 0$ for any $\epsilon$, and clearly $\sum_{k \in N} \phi(q_k) > 0$. Additionally, $\beta\epsilon(1 - \frac{\phi(q_j)}{\phi(q_i)}) > 0$ as $\beta = \gamma - \delta < 0$, and $\phi(q_i) < \phi(q_j)$ as $p_i > p_j$.

To conclude, we have show that the success probability of $q'$ is greater than the success probability of $q$, and the payments are lower, thus the utility of the principal increases when he moves from $q$ to $q'$, which is a contradiction to the optimality of $q$. $\quad\square$

**Observation A.2** *The OR technology exhibits DRS.*

*Proof:* Let $r^a, r^b \in [0,1]^n$ be two profiles of actions, such that $r^b \geq r^a$ (for any $i$, $r^b_i \geq r^a_i$). We need to show that for every $i$, $t_i(r^b_i, r^b_{-i}) - t_i(r^a_i, r^b_{-i}) \leq t_i(r^b_i, r^a_{-i}) - t_i(r^a_i, r^a_{-i})$. Indeed,

$$
\begin{aligned}
t_i(r^b_i, r^b_{-i}) - t_i(r^a_i, r^b_{-i}) &= 1 - (1 - r^b_i)\prod_{j \neq i}(1 - r^b_j) - (1 - (1 - r^a_i)\prod_{j \neq i}(1 - r^b_j)) \\
&= (r^b_i - r^a_i)\prod_{j \neq i}(1 - r^b_j) \\
&\leq (r^b_i - r^a_i)\prod_{j \neq i}(1 - r^a_j) \\
&= 1 - (1 - r^b_i)\prod_{j \neq i}(1 - r^a_j) - (1 - (1 - r^a_i)\prod_{j \neq i}(1 - r^a_j)) \\
&= t_i(r^b_i, r^a_{-i}) - t_i(r^a_i, r^a_{-i})
\end{aligned}
$$

$\quad\square$

# Appendix B. When is Pure Nash Good Enough?

**Lemma B.1** *Let $f : \{0,1\}^n \to \{0,1\}$ for $n \geq 2$ be a monotone Boolean function that is not constant and not a conjunction of some subset of the input bits. Then there exist an assignment to all but two of the bits such that the restricted function is a disjunction of the two bits.*

*Proof:* By induction on the number of bits the function depends on. The base case is $n = 2$, where the only monotone function that is not constant and not a conjunction of some subset of the input bits is the disjunction of two input bits.

Let $x_i$ be a variable on which $f$ depends (which must exist since $f$ is not constant). Let $f^{|x_i=a} = f(a, x_{-i})$ denote the function $f$ restricted to $x_i = a$. We denote $h = f^{|x_i=0}$ and $g = f^{|x_i=1}$. As $f$ is monotone, $f = x \cdot f^{|x_i=1} + f^{|x_i=0} = g \cdot x + h$, where $f^{|x_i=1} \geq f^{|x_i=0}$ (that is, for any $x_{-i}$, if $f(0, x_{-i}) = 1$ then $f(1, x_{-i}) = 1$, and if $f(1, x_{-i}) = 0$ then $f(0, x_{-i}) = 0$). If $h$ is not constant and not a conjunction of some subset of the input bits, then we continue by induction using $h$ by setting $x = 0$. Similarly If $g$ is not constant and not a conjunction of some subset of the input bits, then we continue by induction using $g$ by setting $x = 1$.

So we are left with the case where both $h$ and $g$ are conjunctions of some subset of the variables (where the constant 1 is considered to be the conjunction of the empty set of variables, and it is easy to verify that $h$ and $g$ cannot be the constant 0). Since $f$ depends on $x_i$, we have that $h \neq g$, and since $h \leq g$, there exists some variable $x_j$ ($j \neq i$) that is in the set of variables whose conjunction is $h$ but not in that of $g$. Now set all variables but $x_i$ and $x_j$ to 1, and we are left with $x_i + x_j$. $\square$

**Theorem B.2** *Let $f$ be any monotone Boolean function with $n \geq 2$ inputs, that is not constant and not a conjunction of some subset of the input bits. Then there exist parameters $\{\gamma_i, \delta_i\}_{i=1}^n$ such that the POP of the structured technology with the above parameters (and identical cost $c = 1$) is greater than 1.0233.*

*Proof:* By Lemma B.1 there is an assignment to all but two variables such that the restricted function over the two variables is an OR function. For these two variables we choose the parameters according to the worst POP we know of for an OR technology (see Section 3). For the rest of the variables we choose parameters such that for the value for which the worst utilities ratio is achieved, all the rest of the agents exert no effort and provide success probabilities that are (almost) the success probabilities dictated by the assignment. Next we make this argument formal.

Recall that by Lemma B.1 there is an assignment to all but two variables such that the restricted function over the two variables is an OR function. Let $i_1$ and $i_2$ be the indices of these two variables. In Section 3 we have observed that for OR technology with two agents with values $v = 233$, $\gamma_1 = \gamma_2 = 0.0001$ and $\delta_1 = \delta_2 = 0.9$, the POP is at least 1.0233. We embed this into an instance of an OR technology with $n$ agents by considering a value $v = 233$ and success probabilities as follows: For agents $i_1$ and $i_1$, let $\gamma_{i_1} = \gamma_{i_2} = 0.0001$ and $\delta_{i_1} = \delta_{i_2} = 0.9$. For the rest of the agents, fix a sufficiently small $\epsilon > 0$. Then set $\delta_i = 1 - \epsilon$ and $\gamma_i = 1 - 2\epsilon$ if $i$ was set to 1 in the assignment, and set $\delta_i = 2\epsilon$ and $\gamma_i = \epsilon$ if $i$ was set to 0 in the assignment.

When $\epsilon > 0$ is small enough the payment needed to induce every agent $i \neq i_1, i_2$ to exert effort (for any profile of efforts of the others) will be greater than $v$ as it is inversely proportional to the increase in the success probability due to $i$'s effort, and this goes to zero with $\epsilon$. Thus, for a small enough $\epsilon$ all agents $i \neq i_1, i_2$ will not exert effort in the optimal contract, but each such agent $i$ will provide an almost sure success in the case the assignment of variable $i$ is 1, and an almost sure failure in the case the assignment of variable $i$ was zero. The created technology is essentially the same as the OR technology with agents $i_1$ and $i_2$ with $\gamma_{i_1} = \gamma_{i_2} = 0.0001$, $\delta_{i_1} = \delta_{i_2} = 0.9$, and for the value $v = 233$ the POP will be at least 1.0233. $\square$

# Appendix C. Quantifying the Gain by Mixing

## C.1 POP for $n$ Agents

We observe that for any technology, the POP is bounded by the ratio between the success probability when all agents exert effort, and the success probability when none of the agents exert effort. This simple bound shows that if the success probability when none of the agents exert effort is at least some positive constant, the POP is bounded by a constant.

**Observation C.1** *For any technology $(t, \vec{c})$ with set of agents $N$, $POP(t) \leq \frac{t(N)}{t(\emptyset)}$.*

*Proof:* For any given value $v$, the utility of the principal with the optimal mixed Nash is at most $v \cdot t(N)$, while the utility of the principal with the optimal pure Nash is at least $v \cdot t(\emptyset)$, thus the POP is bounded by $\frac{v \cdot t(N)}{v \cdot t(\emptyset)} = \frac{t(N)}{t(\emptyset)}$. □

From this point we only consider technologies with identical costs. The following lemma shows that anonymous technologies as well as any technology that exhibits DRS have POP at most $n$.

**Lemma C.2** *Assume that for a technology $t$ with $n$ agents the following holds: For any optimal mixed contract $q$ with support $S$, there is a pure profile $a$ with support $T \subseteq S$ such that*

- $t(a) \geq \frac{t(S)}{|S|}$

- *For each agent $i \in T$, and any pure profile $b$ with support $R \subseteq S$ it holds that $t(1, a_{-i}) - t(0, a_{-i}) \geq t(1, b_{-i}) - t(0, b_{-i})$.*

*Then the $POP(t) \leq n$.*

*Proof:* We first observe that $P(a)$, the total payment under the profile $a$ in the case of success, is at most $P(q)$, the total payment under the profile $q$. As $T \subseteq S$, the set of agents that are paid under $a$ is a subset of the set of agents that are paid under $q$. Each agent in $T$ is paid at least as much under $q$, as he is paid under $a$ (by the second condition, as the increase in success probability under $q$ is a convex combination of the increase in success probability for pure profiles with support $R \subseteq S$). Thus, $P(a) \leq P(q)$, and $U(a) > 0$. We conclude that

$$\frac{u(q,v)}{u(a,v)} \leq \frac{t(q)(v - P(q))}{t(a)(v - P(a))} \leq \frac{t(q)}{t(a)} \leq \frac{t(S)}{t(a)} \leq |S|$$

where the last inequality is derived from the first condition. This implies that the POP is bounded by $n$. □

**Corollary C.3** *For any anonymous technology $t$ with $n$ agents, $POP(t) \leq n$.*

*Proof:* Assume that for the value $v$ the mixed profile $q$ is optimal, and its support is of size $k$. Let $t_m$ be the success probability if $m$ agents exert effort, and let $\Delta_m = t_m - t_{m-1}$. Let $m^* = argmax_{m \leq k} \Delta_m$.

By the definition of $m^*$ the second condition holds. The first condition holds as:

$$k \cdot t_m \geq k \cdot (t_0 + t_m - t_0) \geq t_0 + k \cdot (t_m - t_0) \geq t_0 + k \cdot (t_m - t_{m-1}) = t_0 + k \cdot \Delta_m \geq t_0 + (t_k - t_0) = t_k$$

□

**Corollary C.4** *For any technology $t$ with $n$ agents that exhibits DRS and has identical costs, $POP(t) \leq n$.*

*Proof:* Let agent $i \in S$ be the agent with maximal individual contribution in $S$, the support of $q$ ($t(\{i\}) \geq t(\{j\})$ for all $j \in S$). DRS ensures that the two conditions of Lemma C.2 hold. □

The following holds for OR technology with $n$ agents (even non-anonymous), as it exhibits DRS. In particular, even if a single agent has $\delta_i > 1/2$ we get a bound of 2 on the POP.

**Observation C.5** *Assume that the technology $t$ with $n$ agents (with identical costs) exhibits DRS, then $POP(t) \leq \frac{t(N)}{t(\{i\})}$, for agent $i$ with maximal individual contribution ($t(\{i\}) \geq t(\{j\})$ for all $j \in N$).*

*Proof*: Let agent $j \in S$ be the agent with maximal individual contribution in $S$, the support of $q$. Following the proof of Lemma C.2, as $t(\{i\}) \geq t(\{j\})$ and $P(q) \geq P(\{j\}) \geq P(\{i\})$, and $u(q, v) > 0$ ,this implies that $u(\{i\}, v) \geq u(\{j\}, v) > 0$. Thus the optimal pure contract $a^*$ gives utility of at least $u(\{i\}, v) > 0$, therefore for any $v$ we have the bound

$$\frac{u(q, v)}{u(a^*, v)} \leq \frac{u(q, v)}{u(\{i\}, v)} = \frac{t(q)(v - P(q))}{t(\{i\})(v - P(\{i\}))} \leq \frac{t(S)}{t(\{i\})} \leq \frac{t(N)}{t(\{i\})}$$

which implies that the POP is bounded by $\frac{t(N)}{t(\{i\})}$. $\qquad\square$

**Corollary C.6** *For any anonymous technology with $n$ agents that exhibits DRS, it holds that $POP(t) \leq \frac{t_n}{t_1}$.*

## C.2 POP for Anonymous OR

As OR exhibits DRS, the following in a direct corollary of Observation C.5.

**Corollary C.7** *For any anonymous OR technology with $n$ agents, it holds that $POP(OR) \leq \frac{t_n}{t_1}$.*

**Theorem 5.7** *For any anonymous OR technology with $n$ agents:*

1. *If $1 > \delta > \gamma > 0$: (a) $POP \leq \frac{1-(1-\delta)^n}{\delta} \leq n - (n-1)\delta$. (b) POP goes to 1 as $n$ goes to $\infty$ (for any fixed $\delta$) or when $\delta$ goes to 1 (for any fixed $n \geq 2$).*

2. *If $\frac{1}{2} > \gamma = 1 - \delta > 0$: (a) $POP \leq \frac{2(3-2\sqrt{3})}{3(\sqrt{3}-2)}(= 1.154..)$. (b) POP goes to 1 as $\gamma$ goes to 0 or as $\gamma$ goes to $\frac{1}{2}$ (for any fixed $n \geq 2$).*

*Proof*: Based on Corollary C.7, $POP \leq \frac{t(1^n)}{t(1, 0^{n-1})}$, all the results are based on this bound.

1. Proof of part 1(a):

$$\frac{t(1^n)}{t(1, 0^{n-1})} = \frac{1 - (1-\delta)^n}{1 - (1-\delta)(1-\gamma)^{n-1}} \leq \frac{1 - (1-\delta)^n}{1 - (1-\delta)} = \frac{1 - (1-\delta)^n}{\delta}$$

Additionally,

$$\frac{1 - (1-\delta)^n}{1 - (1-\delta)} = \sum_{j=0}^{n-1}(1-\delta)^j = 1 + \sum_{j=1}^{n-1}(1-\delta)^j \leq 1 + \sum_{j=1}^{n-1}(1-\delta) = n - (n-1)\delta$$

and this concludes the proof.

2. Proof of part 1(b):

$$\frac{t(1^n)}{t(1, 0^{n-1})} = \frac{1 - (1-\delta)^n}{1 - (1-\delta)(1-\gamma)^{n-1}}$$

this expression goes to 1 for any fixed $\delta > \gamma > 0$, when $n$ goes to $\infty$, as $(1-\delta)^n$ and $(1-\gamma)^{n-1}$ goes to zero.

Additionally, we saw that $POP \leq \frac{1-(1-\delta)^n}{\delta}$, thus it is clear that if $n$ is fixed and $\delta$ goes to 1, the POP goes to 1.

3. Proof of part 2(a): We first bound the POP for the case of anonymous OR with 2 agents and with $\gamma = 1 - \delta < 1/2$. For this case the POP is bounded by

$$\frac{t(1,1)}{t(0,1)} = \frac{\delta(2-\delta)}{\delta^2 - \delta + 1}$$

The derivative of this ratio is $\frac{2 - 2\delta - \delta^2}{(\delta^2 - \delta + 1)^2}$, which equals to zero at $\delta = \sqrt{3} - 1$. This is a maximum point since the second derivative is negative, and the ratio at this point equals to 1.154... Therefore, $\frac{t(1,1)}{t(1,0)} \leq 1.154...$ Observation C.8 below shows that for any $n \geq 2$ it holds that $\frac{t(1^n)}{t(1,0^{n-1})} \leq \frac{t(1,1)}{t(0,1)}$ thus the same bound holds for any $n$.

4. Proof of part 2(b): The expression $\frac{t(1^n)}{t(1,0^{n-1})} = \frac{1-\gamma^n}{1-\gamma(1-\gamma)^{n-1}}$ goes to 1 when $\gamma$ goes to 0 or $\frac{1}{2}$.

$\square$

For anonymous OR technology with $n$ agents and $\gamma = 1 - \delta < 1/2$ we can bound the POP by 1.154...

**Observation C.8** *Let $OR_{n,\gamma}$ denote the anonymous OR technology of $n$ agents with $\gamma = 1 - \delta < 1/2$. For any $k \geq 3$ it holds that*

$$POP(OR_{k,\gamma}) \leq \frac{OR_{k,\gamma}(1^k)}{OR_{k,\gamma}(1, 0^{k-1})} \leq \frac{OR_{k-1,\gamma}(1^{k-1})}{OR_{k-1,\gamma}(1, 0^{k-2})}$$

*thus for any $k \geq 3$ it holds that*

$$POP(OR_{k,\gamma}) \leq \frac{OR_{k,\gamma}(1^k)}{OR_{k,\gamma}(1, 0^{k-1})} \leq \frac{OR_{2,\gamma}(1,1)}{OR_{2,\gamma}(1,0)} \leq 1.154...$$

*Proof*: For the technology $OR_{k,\gamma}$ it holds that

$$\frac{OR_{k,\gamma}(1^k)}{OR_{k,\gamma}(1, 0^{k-1})} = \frac{1 - \gamma^k}{1 - \gamma \cdot (1-\gamma)^{k-1}}$$

Thus we need to show that for any $k \geq 3$

$$\frac{1 - \gamma^k}{1 - \gamma \cdot (1-\gamma)^{k-1}} \leq \frac{1 - \gamma^{k-1}}{1 - \gamma \cdot (1-\gamma)^{k-2}}$$

which holds if and only if

$$1 - \gamma^k - \gamma \cdot (1 - \gamma)^{k-2} + \gamma^{k+1} \cdot (1 - \gamma)^{k-2} \leq 1 - \gamma^{k-1} - \gamma \cdot (1 - \gamma)^{k-1} + \gamma^k \cdot (1 - \gamma)^{k-1}$$

which holds if and only if

$$-\gamma^{k-1}(1 - \gamma) + \gamma \cdot (1 - \gamma)^{k-2} \cdot (1 - (1 - \gamma)) + \gamma^k \cdot (1 - \gamma)^{k-2} \cdot ((1 - \gamma) - \gamma) \geq 0$$

by dividing by $\gamma^2 \cdot (1 - \gamma)$, this holds if and only if

$$-\gamma^{k-3} + (1 - \gamma)^{k-3} + \gamma^{k-2} \cdot (1 - \gamma)^{k-3}(1 - 2 \cdot \gamma) \geq 0$$

which holds as $1 - \gamma \geq \gamma$ thus $(1 - \gamma)^{k-3} \geq \gamma^{k-3}$ and $\gamma^{k-2} \cdot (1 - \gamma)^{k-3}(1 - 2 \cdot \gamma) \geq 0$.   □

## C.3 POP for 2 Agents

Let us now consider the case that $n = 2$, and prove a better bound on the POP. We have shown that the POP for IRS technology is 1. Since an anonymous technology with 2 agents exhibits either IRS or DRS, we only need to handle the DRS case. Let $\Delta_1 = t_1 - t_0$ and $\Delta_2 = t_2 - t_1$. Assume that $\Delta_1 = \alpha \cdot \Delta_2$ for some $\alpha \geq 1$ (DRS).

The following is a corollary of Lemma 5.3.

**Corollary C.9** *For a DRS technology over 2 agents, assume w.l.o.g. that $t(\{1\}) \geq t(\{2\})$ and denote $\Delta_1 = t(\{1\}) - t(\emptyset)$. For any mixed profile $q = (q_1, q_2)$ it holds that each agent is paid at least $\frac{c}{\Delta_1}$.*

*Proof:* As $t(\{1\}) \geq t(\{2\})$ it implies that $\Delta_1 = t(\{1\}) - t(\emptyset) \geq t(\{2\}) - t(\emptyset)$, and DRS implies that $\Delta_1 = t(\{1\}) - t(\emptyset) \geq t(\{1,2\}) - t(\{1\})$ and $t(\{2\}) - t(\emptyset) \geq t(\{1,2\}) - t(\{2\})$, thus Lemma 5.3 implies that each agent is paid at least $\frac{c}{\Delta_1}$.   □

**Theorem C.10** *For any anonymous technology $t$ with 2 agents, it holds that the $POP(t) \leq 3/2$.*

*Proof:* Let $u((q_1, q_2), v)$ be the utility of the principal for mixed profile $(q_1, q_2)$ when his value for the project is $v$. Let $P(q_1, q_2)$ denote the total payment to both agents if the project is successful. Similarly, let $u((a_1, a_2), v)$ be the utility of the principal for pure profile $(a_1, a_2)$ when his value is $v$.

For a given value $v$, let $(q_1, q_2)$ be the optimal mixed contract, and let $(a_1, a_2)$ be the optimal pure contract. We show that for any value $v$ it holds that $\frac{u((q_1, q_2), v)}{u((a_1, a_2), v)} \leq 3/2$, which is sufficient to prove the theorem.

If the optimal mixed profile is a pure profile, the ratio is 1, thus we only need to handle the case that the profile $(q_1, q_2)$ is not pure (a non-degenerate mixed contract). In this case, as $u((q_1, q_2), v) = t(q_1, q_2) \cdot (v - P(q_1, q_2)) > 0$, it holds that $v - P(q_1, q_2) > 0$. By corollary C.9 this implies that $u((1, 0), v) > 0$ as $P(q_1, q_2) \geq \frac{c}{\Delta_1}$. Thus $u((a_1, a_2), v) \geq u((1, 0), v) > 0$, so

$$\frac{u((q_1, q_2), v)}{u((a_1, a_2), v)} \leq \frac{u((q_1, q_2), v)}{u((1, 0), v)} \leq \frac{t(q_1, q_2)(v - P(q_1, q_2))}{t(1, 0)(v - \frac{c}{\Delta_1})} \leq \frac{t(q_1, q_2)}{t(1, 0)} \leq \frac{t(1, 1)}{t(1, 0)} = \frac{t_2}{t_1}$$

Now we consider two cases. First we consider the case that $t_0 \geq \Delta_2$. In this case

$$\frac{u((q_1,q_2),v)}{u((a_1,a_2),v)} \leq \frac{t_2}{t_1} = \frac{t_0 + \Delta_1 + \Delta_2}{t_0 + \Delta_1} \leq \frac{\Delta_2 + \alpha \cdot \Delta_2 + \Delta_2}{\Delta_2 + \alpha \cdot \Delta_2} = \frac{2 + \alpha}{1 + \alpha} = 1 + \frac{1}{1+\alpha} \leq \frac{3}{2}$$

to replace $t_0$ with $\Delta_2$ we use Lemma C.13.

Next we consider the case that $t_0 < \Delta_2$. In this case we look at the value $v^*$ for which the principal is independent between contracting with 1 or 2 agents. At $v = v^*$ it holds that $t(1,0) \cdot (v - \frac{c}{\Delta_1}) = t(1,1) \cdot (v - \frac{2c}{\Delta_2})$, thus $v \cdot \Delta_2 = v \cdot (t_2 - t_1) = t_2 \frac{2c}{\Delta_2} - t_1 \cdot \frac{c}{\alpha \cdot \Delta_2}$, thus it holds that $v^* = \frac{c}{\alpha(\Delta_2)^2}(2\alpha \cdot t_2 - t_1)$. For a value $v \leq v^*$ it is enough to bound the ratio $\frac{u((q_1,q_2),v)}{u((1,0),v)}$, while for a value $v \geq v^*$ it is enough to bound the ratio $\frac{u((q_1,q_2),v)}{u((1,1),v)}$. We bound each of these ratios separately.

By Lemma C.13, for the case that $0 \leq t_0 < \Delta_2$, $\frac{t_2}{t_1} = \frac{t_0 + \Delta_1 + \Delta_2}{t_0 + \Delta_1} \leq \frac{(1+\alpha)\Delta_2}{\alpha \cdot \Delta_2} = 1 + \frac{1}{\alpha}$.
For a value $v \leq v^*$

$$\frac{u((q_1,q_2),v)}{u((1,0),v)} \leq \frac{t(q_1,q_2)(v - P(q_1,q_2))}{t(1,0)(v - \frac{c}{\Delta_1})} \leq \frac{t_2}{t_1} \cdot \frac{v - P(q_1,q_2)}{v - \frac{c}{\Delta_1}} \leq \left(1 + \frac{1}{\alpha}\right) \cdot \frac{v - \frac{2c}{\Delta_1}}{v - \frac{c}{\Delta_1}} \leq$$

$$\left(1 + \frac{1}{\alpha}\right) \cdot \left(1 - \frac{1}{\frac{v^*}{c} \cdot \Delta_1 - 1}\right)$$

Now, as $\frac{\Delta_2}{t_2} = \frac{\Delta_2}{t_0 + (1+\alpha)\Delta_2} \geq \frac{\Delta_2}{\Delta_2 + (1+\alpha)\Delta_2} = \frac{1}{2+\alpha}$, we conclude that

$$\frac{1}{\frac{v^*}{c} \cdot \Delta_1 - 1} = \frac{1}{\frac{\Delta_1}{\alpha(\Delta_2)^2}(2\alpha \cdot t_2 - t_1) - 1} = \frac{\Delta_2}{2\alpha \cdot t_2 - t_1 - \Delta_2} = \frac{\Delta_2}{(2\alpha - 1) \cdot t_2} \geq$$

$$\frac{1}{(2\alpha - 1)(2 + \alpha)}$$

Thus

$$\frac{u((q_1,q_2),v)}{u((1,0),v)} \leq \left(1 + \frac{1}{\alpha}\right) \cdot \left(1 - \frac{1}{\frac{v^*}{c} \cdot \Delta_1 - 1}\right) \leq \left(1 + \frac{1}{\alpha}\right) \cdot \left(1 - \frac{1}{(2\alpha - 1)(2 + \alpha)}\right)$$

Lemma C.11 shows that the function on the RHS is bounded by $3/2$ for any $\alpha \geq 1$.
Finally, for a value $v \geq v^*$, it is enough to bound the ratio $\frac{u((q_1,q_2),v)}{u((1,1),v)}$.

$$\frac{u((q_1,q_2),v)}{u((1,1),v)} = \frac{t(q_1,q_2)(v - P(q_1,q_2))}{t(1,1)(v - \frac{2c}{\Delta_2})} \leq \frac{v - \frac{2c}{\Delta_1}}{v - \frac{2c}{\Delta_2}} = \frac{v - \frac{2c}{\alpha \cdot \Delta_2}}{v - \frac{2c}{\Delta_2}}$$

Intuitively, as the fraction goes to 1 as $\alpha$ goes to 1, this implies that for sufficiently small $\alpha$ the fraction is less than $3/2$. Formally,

$$\frac{v - \frac{2c}{\alpha \cdot \Delta_2}}{v - \frac{2c}{\Delta_2}} = 1 + \frac{\frac{2c}{\Delta_2} - \frac{2c}{\alpha\Delta_2}}{v - \frac{2c}{\Delta_2}} = 1 + \frac{2(1 - \frac{1}{\alpha})}{\frac{v}{c} \cdot \Delta_2 - 2} \leq 1 + 2\left(\frac{\alpha - 1}{\alpha}\right) \cdot \frac{1}{\frac{v^*}{c} \cdot \Delta_2 - 2} \leq$$

$$1 + 2\left(\frac{\alpha - 1}{\alpha}\right) \cdot \frac{1}{\frac{\Delta_2}{\alpha(\Delta_2)^2}(2\alpha \cdot t_2 - t_1) - 2} \leq 1 + \frac{2(\alpha - 1)\Delta_2}{2\alpha \cdot t_2 - t_1 - 2\alpha \cdot \Delta_2} = 1 + \frac{2(\alpha - 1)\Delta_2}{(2\alpha - 1)t_1} \leq$$

$$1 + \frac{2(\alpha - 1)}{(2\alpha - 1)\alpha}$$

We find the maximum of the RHS. The derivative by $\alpha$ of $1 + \frac{2(\alpha-1)}{(2\alpha-1)\alpha}$ is $\frac{-2(2 \cdot \alpha^2 - 4 \cdot \alpha + 1)}{(2 \cdot \alpha - 1)^2 \cdot \alpha^2}$. The maximum is obtained at $\alpha = 1 + \frac{\sqrt{2}}{2}$ (it is a maximum as the second derivative is negative for $\alpha = 1 + \frac{\sqrt{2}}{2}$), and the maximum is $1 + \frac{\sqrt{2}}{(1+\sqrt{2}) \cdot (1 + \frac{\sqrt{2}}{2})} < 1.35$. □

**Lemma C.11** *For $\alpha \geq 1$ it holds that*

$$\left(1 + \frac{1}{\alpha}\right) \cdot \left(1 - \frac{1}{(2\alpha - 1)(2 + \alpha)}\right) \leq 3/2$$

*Proof*:

$$f(\alpha) = \left(1 + \frac{1}{\alpha}\right) \cdot \left(1 - \frac{1}{(2\alpha - 1)(2 + \alpha)}\right) = \left(1 - \frac{1}{\alpha + 2}\right) \cdot \frac{2\alpha^2 + 3\alpha - 3}{\alpha(2\alpha - 1)}$$

Let $h(\alpha) = \frac{2\alpha^2 + 3\alpha - 3}{\alpha(2\alpha - 1)}$. The derivative of $h(\alpha)$ by $\alpha$ is $\frac{-8\alpha^2 + 12\alpha - 3}{\alpha^2(2\alpha - 1)^2}$, it has a maximum at $\frac{3 + \sqrt{3}}{4}$, and the value of the maximum is lower than $2.072$.

We look at $\tilde{\alpha} = 8/5$. As the function $1 - \frac{1}{\alpha + 2}$ is an increasing function of $\alpha$ (for $\alpha \geq 1$), we get that for any $\alpha \leq \tilde{\alpha}$, $f(\alpha) \leq \left(1 - \frac{1}{\tilde{\alpha} + 2}\right) \cdot 2.072 = \frac{13}{18} * 2.072 < 3/2$ To conclude the proof we show that $f(\alpha)$ is a decreasing function of $\alpha$, for any $\alpha \geq \tilde{\alpha} = 8/5$. The derivative of $f(\alpha)$ by $\alpha$ is

$$\frac{-2(2\alpha^4 + 4\alpha^3 - 4\alpha^2 - 9\alpha + 3)}{\alpha^2(2\alpha - 1)^2(2 + \alpha)^2}$$

Thus we only need to show that $2\alpha^4 + 4\alpha^3 - 4\alpha^2 - 9\alpha + 3 > 0$ for any $\alpha \geq \tilde{\alpha} = 8/5$. This holds as $4\alpha^3 - 4\alpha^2 = 4\alpha^2(\alpha - 1) > 0$ for any $\alpha > 1$, and $2\alpha^4 - 9\alpha + 3 \geq 2\tilde{\alpha}^4 - 9\tilde{\alpha} + 3 = \frac{1067}{625} > 0$ for any $\alpha \geq \tilde{\alpha}$ (as the function $2\alpha^4 - 9\alpha + 3$ has derivative $8\alpha^3 - 9$ which is positive for any $\alpha \geq 9^{1/3}/2$, thus it is a monotonically increasing function for $\alpha \geq \tilde{\alpha} = 8/5 > 1.05 > 9^{1/3}/2$ ). □

**Theorem C.12** *For any technology $t$ (even non-anonymous) with 2 agents and identical costs, it holds that the $POP(t) \leq 2$.*

*Proof*: If the technology is anonymous, we have already proved a stronger claim. Assume that it is not, then w.l.o.g. assume that $t(1,0) \geq t(0,1)$. We have shown that the profile $(0,1)$ is never optimal, this implies (by the same argument as we have seen in the case that the technology is anonymous), that

$$POP \leq \frac{u((q_1, q_2), v)}{u((a_1, a_2), v)} \leq \frac{u((q_1, q_2), v)}{u((1,0), v)} \leq \frac{t(1,1)}{t(1,0)}$$

If the technology exhibits IRS, then we know that POP=1. To conclude the proof we show that if the technology does not exhibits IRS then $\frac{t(1,1)}{t(1,0)} \leq 2$. Assume that $\frac{t(1,1)}{t(1,0)} > 2$, we show that the technology exhibits IRS. This is true since $\frac{t(1,1)}{t(1,0)} > 2$ implies:

$$t(1,1) - t(1,0) > t(1,0) \geq t(1,0) - t(0,0)$$

and as $t(1, 0 > t(0, 1)$ it also holds

$$t(1, 1) - t(0, 1) > t(1, 1) - t(1, 0) \geq t(1, 0) \geq t(0, 1) \geq t(0, 1) - t(0, 0)$$

which implies IRS. □

**Lemma C.13** *If $a \geq b \geq 0$ and $x \geq y > 0$ then $\frac{a+x}{a+y} \leq \frac{b+x}{b+y}$.*

# Appendix D. The Robustness of Mixed Nash Equilibria

**Theorem 6.2** *If the mixed optimal contract $q$ includes at least two agents that truly mix ($\exists i, j$ s.t. $q_i, q_j \in (0, 1)$), then $q$ is not a strong equilibrium.*

*Proof:* Let $Q$ be the support of $q$ (i.e., $Q = \{i | q_i > 0\}$), and let $k = |Q|$. Recall that the optimal payments that induce the strategy profile $q$ are $p_i = \frac{c_i}{\Delta_i(q_{-i})}$ (where $\Delta_i(q_{-i}) = t(1, q_{-i}) - t(0, q_{-i})$) for any $i \in Q$, and $p_i = 0$ for any $i \in N \setminus Q$. Let $\Gamma = \{i | q_i \in (0, 1)\}$ ($|\Gamma| \geq 2$ by assumption), and consider a deviation of the coalition $\Gamma$ into a pure strategy profile $q'_\Gamma$, in which for any $i \in \Gamma$, $q_i = 1$. $q'$ denote the new profile (i.e., $q' = (q'_\Gamma, q_{-\Gamma})$).

We next show that for any $i \in \Gamma$, $u_i(q) < u_i(q')$, thus $q$ is not resilient to a deviation by $\Gamma$. Since $q_i \in (0, 1)$, $i$ must be indifferent between $a_i = 0$ and $a_i = 1$ (see claim 2.1); therefore, $i$'s utility in $q$ is:

$$u_i(q) = u_i(0, q_{-i}) = c_i \frac{t(0, q_{-i})}{\Delta_i(q_{-i})}$$

The utility of $i$ in $q'$ is:

$$u_i(q') = t(q')p_i - c_i = t(q') \frac{c_i}{\Delta_i(q_{-i})} - c_i = c_i \left( \frac{t(q')}{\Delta_i(q_{-i})} - 1 \right) = c_i \left( \frac{t(q') - t(1, q_{-i}) + t(0, q_{-i})}{\Delta_i(q_{-i})} \right)$$

Therefore, $u_i(q') > u_i(q)$ if and only if $t(q') - t(1, q_{-i}) > 0$, which holds by the assumption that $|\Gamma| \geq 2$ and the monotonicity of $t$. □

# Appendix E. Algorithmic Aspects

## E.1 Results for the Mixed Case

We next show that in the black box model, exponential number of queries is needed to determine the optimal mixed contract. We have proved this for the optimal pure contract (for completeness we present the claim as Theorem E.2, taken from (Babaioff et al., 2006a)), and now show that it also holds for the mixed case.

**Theorem 7.1** *Given as input a black box for a success function $t$ (when the costs are identical), and a value $v$, the number of queries that is needed, in the worst case, to find the optimal mixed contract is exponential in $n$.*

*Proof:* We show that the optimal mixed contract for the technology presented in Theorem E.2 at the value $c(k + 1/2)$ has support exactly $\hat{T}$, thus the claim is a direct result of Theorem E.2.

Assume that $q$ is optimal mixed contract at the value $c(k+1/2)$. The support of $q$ must be of size at most $k$, otherwise the payment in case of success is at least $c(k+1) > c(k+1/2)$ (as each agent in the support must be paid at least $c$), which implies negative utility. If he support is of size at most $k$ and is not exactly $\hat{T}$, then there is at least one agent that is paid $c/\epsilon > c(k+1/2)$ for sufficiently small $\epsilon > 0$. Thus in any such case the utility is again negative. □

Next we show that for read-one network the optimal mixed contract is #$P$-hard. It is based on a theorem from the work of Babaioff et al. (2006a) cited as Theorem E.3 below.

**Theorem 7.2** *The Optimal Mixed Contract Problem for Read Once Networks is #$P$-hard (under Turing reductions).*

*Proof:* We use the reduction presented in Theorem E.3. We prove that for $\gamma_x$ close enough to $1/2$, at the transition point from $E$ to $E \cup \{x\}$ of the pure case, the optimal mixed contract is pure (also $E$ and $E \cup \{x\}$). This implies that we can use the same argument of Theorem E.3 to calculate the network reliability (which is #$P$-hard) using an algorithm for the optimal mixed contract.

Lemma E.1 below presents a generalization of a lemma from the work of Babaioff et al. (2006a) to the mixed case. The lemma implies that at the value $v$ that $x$ is first entered to the support of the optimal mixed contract $q$, the contract for $x$ is optimal for the value $v \cdot t(E)$. But for a single edge, the only optimal mixed contracts are pure, thus $x$ exerts effort with probability 1. Additionally, the contract for the original graph (with edges $E$) is optimal for the value $v \cdot (1 - \gamma_x)$, thus for $\gamma_x$ close enough to $1/2$, $v$ is large enough such that the optimal mixed contract is with all agents exerting effort with probability 1 (pure). □

Let $g$ and $h$ be two Boolean functions on disjoint inputs and let $f = g \bigwedge h$ (i.e., take their networks in series). The optimal mixed contract for $f$ for some $v$, denoted by $q_S$, is composed of the $h$-part and the $g$-part, call them mixed profile for these parts $q_T$ and $q_R$ respectively.

**Lemma E.1** *Let $q_S$ be an optimal mixed contract for $f = g \bigwedge h$ on $v$. Then, $q_T$ is an optimal mixed contract for $h$ on $v \cdot t_g(q_R)$, and $q_R$ is an optimal mixed contract for $g$ on $v \cdot t_h(q_T)$.*

The proof is the same as the proof for the pure case, presented in the work of Babaioff et al. (2006b).

## E.2 Results from the work of Babaioff et al. (2006b) for the Pure Case

The following results are cited from the work of Babaioff et al. (2006b), for completeness.

**Theorem E.2** *Given as input a black box for a success function $t$ (when the costs are identical), and a value $v$, the number of queries that is needed, in the worst case, to find the optimal contract is exponential in $n$.*

*Proof:* Consider the following family of technologies. For some small $\epsilon > 0$ and $k = \lceil n/2 \rceil$ we define the success probability for a given set $T$ as follows. If $|T| < k$, then $t(T) = |T| \cdot \epsilon$. If $|T| > k$, then $t(T) = 1 - (n - |T|) \cdot \epsilon$. For each set of agents $\hat{T}$ of size $k$, the technology $t_{\hat{T}}$ is defined by $t(\hat{T}) = 1 - (n - |\hat{T}|) \cdot \epsilon$ and $t(T) = |T| \cdot \epsilon$ for any $T \neq \hat{T}$ of size $k$.

For the value $v = c \cdot (k + 1/2)$, the optimal contract for $t_{\hat{T}}$ is $\hat{T}$ (for the contract $\hat{T}$ the utility of the principal is about $v - c \cdot k = 1/2 \cdot c > 0$, while for any other contract the utility is negative).

If the algorithm queries about at most $\binom{n}{\lceil n/2 \rceil} - 2$ sets of size $k$, then it cannot always determine the optimal contract (as any of the sets that it has not queried about might be the optimal one). We conclude that $\binom{n}{\lceil n/2 \rceil} - 1$ queries are needed to determine the optimal contract, and this is exponential in $n$. □

Let $t(E)$ denote the probability of success when each edge succeeds with probability $\delta_e$. We first notice that even computing the value $t(E)$ is a hard problem: it is called the network reliability problem and is known to be $\#P - hard$ (Provan & Ball, 1983). Just a little effort will reveal that our problem is not easier:

**Theorem E.3** *The Optimal Contract Problem for Read Once Networks is $\#P$-hard (under Turing reductions).*

*Proof*: We will show that an algorithm for this problem can be used to solve the network reliability problem. Given an instance of a network reliability problem $< G, \{\zeta_e\}_{e \in E} >$ (where $\zeta_e$ denotes $e$'s probability of success), we define an instance of the optimal contract problem as follows: first define a new graph $G'$ which is obtained by "And"ing $G$ with a new player $x$, with $\gamma_x$ very close to $\frac{1}{2}$ and $\delta_x = 1 - \gamma_x$. For the other edges, we let $\delta_e = \zeta_e$ and $\gamma_e = \zeta_e/2$. By choosing $\gamma_x$ close enough to $\frac{1}{2}$, we can make sure that player $x$ will enter the optimal contract only for very large values of $v$, after all other agents are contracted (if we can find the optimal contract for any value, it is easy to find a value for which in the original network the optimal contract is $E$, by keep doubling the value and asking for the optimal contract. Once we find such a value, we choose $\gamma_x$ s.t. $\frac{c}{1 - 2\gamma_x}$ is larger than that value). Let us denote $\beta_x = 1 - 2\gamma_x$.

The critical value of $v$ where player $x$ enters the optimal contract of $G'$, can be found using binary search over the algorithm that supposedly finds the optimal contract for any network and any value. Note that at this critical value $v$, the principal is indifferent between the set $E$ and $E \cup \{x\}$. Now when we write the expression for this indifference, in terms of $t(E)$ and $\Delta_i^t(E)$ , we observe the following.

$$t(E) \cdot \gamma_x \cdot \left( v - \sum_{i \in E} \frac{c}{\gamma_x \cdot \Delta_i^t(E \setminus i)} \right) = t(E)(1 - \gamma_x) \left( v - \sum_{i \in E} \frac{c}{(1 - \gamma_x) \cdot \Delta_i^t(E \setminus i)} - \frac{c}{t(E) \cdot \beta_x} \right)$$

if and only if

$$t(E) = \frac{(1 - \gamma_x) \cdot c}{(\beta_x)^2 \cdot v}$$

thus, if we can always find the optimal contract we are also able to compute the value of $t(E)$. □